\documentclass[11pt]{article}

\usepackage{graphicx,pstricks}
\usepackage{moreverb}
\usepackage{subfigure}
\usepackage{amsmath}
\usepackage{mathrsfs}
\usepackage{epsfig}
\usepackage{txfonts}
\usepackage{palatino}
\usepackage{epstopdf}
\usepackage{xfrac}
\usepackage{mdwlist}
\usepackage{morefloats}
\usepackage{cite}
\usepackage{multirow}
\usepackage{authblk}

\DeclareMathAlphabet{\mathsfsl}{OT1}{cmss}{m}{sl}

\DeclareMathOperator{\trace}{tr}

\newcommand{\mb}[1]{\mbox{\boldmath $#1$}}
\newcommand{\beqn}{\begin{equation}}
\newcommand{\eeqn}{\end{equation}}

\newcommand{\fepx}{{\bfseries{\slshape{FEpX}}}}

\newcommand{\vctr}[1]{\boldsymbol{#1}}

\newcommand{\tnsr}[1]{\mathsfsl{#1 }}

\newcommand{\pddefrate}{{\mb{D}^\prime}^p}

\newcommand{\cauchy}{{\boldsymbol{\sigma}}}
\newcommand{\dcauchy}{{\boldsymbol{\sigma}}^\prime}

\newcommand{\schmid}{{\hat{\vctr{s}}}^{\alpha}\otimes {\hat{\vctr{m}}}^{\alpha}}
\newcommand{\symschmid}{\hat{\tnsr{p}}^\alpha}

\newcommand{\rss}{\tau^\alpha}
\newcommand{\gammadot}{\dot{\gamma}^\alpha}
\newcommand{\gdotz}{\dot{\gamma}_0}

\newcommand{\sumss}{\sum_{\alpha}}

\begin{document}
\title{A methodology to evaluate continuum-scale yield surfaces\\ based on the spatial distributions of yielding at the crystal scale\\ 
}
\author[1]{Andrew C. Poshadel}
\author[1]{Paul R. Dawson}
\affil[1]{Sibley School of Mechanical and Aerospace Engineering, Cornell University, Ithaca, New York, USA}
\date{}
\maketitle

\begin{abstract}
Using a correlation between local yielding and a multiaxial strength-to-stiffness parameter, the continuum-scale yield surface for a polyphase, polycrystalline solid is predicted.  The predicted surface explicitly accounts for microstructure through the quantification of strength-to-stiffness based on a finite element model of a crystal-scale sample.  The multiaxial strength-to-stiffness is evaluated from the elastic response of the sample and the restricted slip, single-crystal yield surface.  Macroscopic yielding is defined by the propogation of a yield band through the sample and is detected with the aid of a flood fill agorithm.   The methodology is demonstrated with the evaluation of a plane-stress yield surface for a dual-phase superaustenitic stainless steel.

\end{abstract}
\clearpage

\section{Introduction}
\label{sec:introduction}

A physically-based methodology  that explicitly incorporates a polycrystalline solid's microstructure is presented for calculating its macroscopic yield surface.  The methodology analyzes the  spatial distribution of the onset of yielding at the microscale. 
This distribution is predicted using a correlation between the multiaxial strength-to-stiffness ratio and the likelihood of a crystal yielding locally that has been reported previously in  \cite{pos_daw_multiaxial-y2e} and \cite{pos_daw_twophase}.  
The multiaxial strength-to-stiffness ratio is determined from the elastic response of
a virtual sample of the solid, together with knowledge of the single crystal yield surface.
A macroscopic yield detection algorithm is employed to assess if the spatial distribution of
yielding at the crystal scale is sufficient to render yielding of the full sample macroscopically.

 To demonstrate the new methodology, the yield surface is evaluated for a dual-phase, stainless steel alloy, LDX-2101. 
 This alloy has nearly equal volume fractions of its austenitic (face-centered cubic) and ferrite (body-center cubic) phases.
 Multiple, biaxial stress states are examined and used to construct a plane stress yield surface in deviatoric stress space.
For several different biaxial stress states,  predicted yield stress is compared to one constructed using elasto-plastic finite element simulation data for loadings that extend past the yield point and induce plastic flow over the entire sample.
The influence of the relative values of the slip system strengths between the ferritic and austenitic phases 
on the macroscopic yield surface is illustrated by constructing a second yield surface assuming different crystal properties for the two phases.  In particular, phase strengths differing by a factor of two, instead of being equal, were assumed (in both cases, the alloys have the same volume-averaged strength). 
The example effectively illustrates that the new methodology provides an efficient approach for  building  a continuum-scale yield surface that explicitly accounts for microstructural structural features like phase properties, phase topology, grain morphology, and crystallographic texture.

The paper is organized as follows.  First, yield surfaces at the single-crystal and the continuum scales are briefly summarized.  Next, an overview of the complete process of evaluating the macroscopic yield surface is presented, along with summaries of the main components of the method.  This includes the multiaxial strength-to-stiffness parameter, the local yield predictor, and the algorithm for detecting macroscopic yield in a sample with a distribution of local yielding.  Following these summaries, the finite element model of a  sample of LDX-2101 is discussed, including a brief review of the principal points of the finite element framework employed here, the virtual sample instantiation,
the assignment of local properties, and the suite of simulations conducted. 
Next, the spatial distribution of yielding over a virtual polycrystal is discussed for the case of uniaxial tension to illustrate the coupling between the microstructure and the the onset of yielding.  The demonstration of the
method for evaluating the macroscopic yield surface for LDX-2101 is then presented. Following this, the impact of changing the relative values of slip system strength of the yield surface is used to show how the microstructure influences the macroscopic yield surface.  
 Finally, conclusions are offered.



\section{Yield criteria for anisotropic, polycrystalline solids}
\label{sec:background}
The intent of the methodology presented in this paper is to determine the macroscopic yield surface
given knowledge of the single crystal yield surface(s) and of the microstructure.  
The methodology is applicable to polyphase, polycrystalline solids provided the single-crystal yield surfaces are known for all of the structurally significant phases.  In this section the construct 
of the single-crystal yield surface for crystal deforming by restricted slip (slip on a limited number of systems) is summarized.  This is followed by an analogous summary of continuum yield surfaces, including two common approaches for representing the surfaces mathematically: analytical and piecewise functions.
A more thorough review of anisotropic yield criteria that account for  crystallographic effects is
available in an article appearing in {\bf International Material Reviews}~\cite{Dawson03a}.

Criteria at both scales share the characteristic that they form closed surfaces in deviatoric stress space that define the stress states capable of inducing plastic flow.  No plasticity occurs for stress states within the surfaces; stress states outside the surfaces are inadmissable. 
Rate dependence of plastic flow alters this picture, but is accommodated by designating a family of (flow) surfaces parameterized by a metric that quantifies the kinetics of slip.  
One value of the metric, defined by a particular combination of strain rate and temperature, can be designated as a reference.

A critical point to note is that the methodology presented in this paper provides yield stress data, generated numerically using the crystal-scale properties for slip within crystals, from which the controlling parameters for any one of the continuum yield surface representations could be determined.  In the demonstration presented later, a piecewise representation is used as a matter of the authors' preference, but an analytical form would be possible as well.

\subsection{Yield surfaces for single crystals}
\label{sec:scys}
For crystalline solids loaded at low to moderate temperature and moderate strain rates, the plastic deformation within a crystal often is dominated by crystallographic slip on a limited number of slip systems (commonly called restricted slip). Plastic flow by crystallographic slip is assumed to be isochoric and independent of the mean stress. It is also highly nonlinear and rate-dependent (viscoplastic). 

The yield surface for a single crystal (SCYS) is a combination of the separate relations for individual slip systems that each designate the stress required to activate slip on the system. For each slip system, $k$, the relation is that the magnitude of the resolved shear stress equals a critical value, where the resolved shear stress, $\rss$, is the projection of  the deviatoric crystal stress tensor, $\dcauchy$, onto the slip plane and into the slip direction.\footnote{For ease of reading here, the relations are written in the current configuration without making distinctions between various configurations that arise in a kinematically nonlinear framework. Readers are referred to \cite{Dawson14a} for more complete exposition.} This  condition is simply stated using the symmetric portion of the Schmid tensor, $\symschmid$, as:
 \begin{equation}
\rss=\trace( \symschmid  \dcauchy) = \pm \tau^{\rm cr}
\label{eq:rss_projection}
\end{equation}
where the Schmid tensor for each slip system is defined by the diadic product of the slip plane normal, $\vctr{m}^k$, and slip direction, $\vctr{s}^k$. That is:
\begin{equation}
\symschmid  =  \mbox{\rm sym}(\schmid) 
\end{equation}
The critical value of the resolved shear stress, $\tau^{\rm cr}$, is rate dependent as a consequence of the kinetics of slip. The influence of the slip kinetics may be approximated mathematically over a modest range of strain rates using a power law expression between the resolved shear stress and the slip system shearing rate, $\gammadot$:
\begin{equation}
\tau^{\rm cr}=g^{k}({\frac{\| \gammadot \|}{\gdotz}})^m 
\label{eq:ss_kinetics}
\end{equation}
where $m$ is the rate sensitivity and $\gdotz$ is a scaling parameter. 
The critical resolved shear stress is scaled by the slip system strength, $g^k$, which in general may be different for the various systems.

These relations collectively define a set of polytopes in deviatoric (five-dimensional) stress space, parameterized by the crystal plastic deformation rate, from the inner envelope of the set of planes given by Equation~\ref{eq:rss_projection}~\cite{backofen_book,kocks_book}. If the crystal stress lies on a face of the yield surface, the crystal is plastically deforming via slip on the slip system associated with that face. For polyslip (plastic deformation involving multiple slip systems), the stress lies on an edge at the intersection of two faces or in a vertex created by the intersections of two or more edges. The deviatoric crystal plastic deformation rate is a linear combination of the slip system rates: 
\begin{equation}
{\tnsr{d}^p}^\prime =  \sumss \gammadot \symschmid
\label{eq:plasdefrate}
\end{equation}
where the symmetric portion of the Schmid tensor defines the simple shear deformation mode associated with deformation by slip for each system.
\subsection{Yield surfaces for continua}
At the continuum scale, specific microstructural information does not appear in the yield criteria.  Rather, the criteria are functions of the stress together with scaling based on a metric 
for the strength.   Anisotropic criteria emerged as extensions of those proposed by \cite{tresca}, \cite{huber}, 
\cite{von_mises}, and \cite{hencky} 
and recover one of these in the limit of isotropic response.
For metals that exhibit no inelastic volumetric deformation,
a yield criterion is a convex surface in deviatoric stress space:  
\beqn
f(\mb{\sigma^\prime}) =  0
\label{eq:macro_ys}
\eeqn
where $\cauchy^\prime$ is the deviatoric part of the Cauchy stress~\cite{gurtin_book}.
Anisotropy implies that the surface given by $f$ cannot be a function solely of the invariants of the stress.  
Rather, the parameters in the criteria are linked to a specific coordinate
basis attached to the material. 
Rate dependence may be 
parameterized by a variable such as the Zener-Holloman parameter or the Fisher factor, $\cal F$:
\beqn
f(\mb{\sigma^\prime}, {\cal F}) = 0 .
\label{eq:visco_ys}
\eeqn
With this representation of the inelastic behavior, plastic flow occurs at any stress level,
but because of the highly nonlinear dependence of the flow surface on the
rate parameter, the plastic deformation rate drops off
rapidly as the stress diminishes. 
The plastic deformation strain rate may be derived from the yield criterion if the criterion is
constructed as a potential function.  
In this case, the plastic deformation, $\pddefrate$ , rate is:  
\beqn
\pddefrate = \frac{\partial f}{\partial \dcauchy} 
\eeqn
and $f$ is proportional to the plastic rate of work.

\subsubsection{ Yield surfaces defined by analytical functions in stress space}
One of the earliest analytical criteria for aniostropic behavior is one proposed by Hill~\cite{hill_48}:
\beqn
2 f(\mb{\sigma}^\prime) =  
c_1{(\sigma_{22} -\sigma_{33})}^2 +
c_2{(\sigma_{33} -\sigma_{11})}^2 +
c_3{(\sigma_{11} -\sigma_{22})}^2 +
2c_4\sigma_{23}^2+
2c_5\sigma_{31}^2+
2c_6\sigma_{12}^2
-1
\label{eq:hill-48}
\eeqn
where $c_1 - c_6$ are parameters evaluated from experimental data.
This criterion, while being widely applied, is seen as too inflexible to describe 
yield surfaces for metals generally, regardless of whether they have cubic or non-cubic crystal structures. 
This assessment stems in part from what is referred to as its anomalous behavior with respect to the
combination of stress and strain direction predicted by the criterion.
Efforts to improve upon Equation~\ref{eq:hill-48} are many, with progress having been 
reported on:  
(1) the development of forms applicable to special stress states (particularly plane stress); 
(2) the exponential power on the stress (which in large measure dictates the sharpness of the corners);
(3) the stress transformation (which gives weighting of components and directionality with
respect to the material orientation); and, 
(4) the treatment of strain hardening (for expansion of the flow surface during deformation).
Two criteria that have been adopted widely are those developed by Hill~\cite{Hill90a} and  by Barlat et al.~\cite{Barlat97a} 
A useful summary is given by Hosford (see Table 8.1 of \cite{hosford_book}).

\subsubsection{ Yield surfaces defined by piecewise functions in stress space}
\label{sec:piecewise-reps}
\leftline{\it Hyperplanes}  
To account for crystallographic texture, Maudlin and coworkers~\cite{mau_wri_gra_hou_95p}
have developed a methodology that utilizes a flow surface 
defined by hyperplanes in deviatoric stress space. 
The shape of the flow surface is defined by the inner envelope of
planes:
\beqn
 f_\chi \equiv \zeta_{ij}^\chi \sigma_{ij} - Y^\chi = 0 \hspace{1cm} \chi = 1, 2, ..., m 
\label{eq:maudlin_ys}
\eeqn
and then scaled to the appropriate size according to:
\beqn
\sigma_{ij} = \tilde \sigma_{ij} \frac{Y}{\bar M} 
\hspace{1cm}{\rm and} \hspace{1cm}
Y^\chi = {\tilde Y}^\chi  \frac{ Y}{\bar M}.
\eeqn
 Here, $\tilde \sigma_{ij}$ is a unit-magnitude stress direction, $\bar M$ is the average Taylor factor~\cite{kocks_book},
$Y$ is an effective flow stress that provides an overall
size to the surface,
$Y^\chi$ is the minimum distance from the origin of
deviatoric stress space to the $\chi$ hyperplane, and
$\zeta_{ij}^\chi$ are the coefficients of the
normal vector to the $\chi$ hyperplane.  
The hyperplane data is determined by probing of 
the flow surface using crystal plasticity, together with the
measured crystallographic texture for a material, to find the flow strength
in a number of stress directions.  This set of points
is used in a tesselation algorithm to determine a
piecewise surface defined by Equation~\ref{eq:maudlin_ys}.   
Also needed is the flow law, which is obtained by
taking $f_\chi$ as a plastic potential and differentiating with
respect to the stress:
\beqn
\dot\epsilon^p_{ij} = 
\sum_{\chi = 1}^{m_{act}}  \dot\lambda^\chi \zeta_{ij}^\chi 
\label{eq:maudlin_fl}
\eeqn
where the linearity of $f_\chi$ with respect to $\sigma_{ij}$ has been 
exploited, $\dot\lambda^\chi$ are referred to as the loading scalars, and 
$m_{act}$ is the number of active hyperplanes.

\leftline{\it Piecewise functions}  Yield surfaces also can be represented using piecewise interpolants
({\it e.g.} finite elements) in stress stress space~\cite{daw_boy_hal_dur_04}.
The yield surface is a polytope in stress space tiled by elements that define the surface.  
Kinetics of plastic flow (rate sensitivity) is accommodated by defining a reference surface for a prescribed stain rate and temperature combination, as suggested in Equation~\ref{eq:visco_ys}:
\beqn
\dcauchy_{ref} = \left[ N(\mb{\xi}) \right] 
\left\{ \dcauchy_j  \right\}
\eeqn
where 
$\dcauchy_{ref}$ is the value of the stress at yielding at the reference value of  $\cal F$, $N(\mb{\xi})$ represents a set of continuous piecewise interpolation functions,
and $\mb{\xi}$ are local coordinates on a four dimensional surface used in the interpolation.
The nodal point values of the stress at yield, $\left\{ \dcauchy_j  \right\}$,
are evaluated from data, such as with the method proposed here, for the reference
combination of strain rate and temperature.
Values of the yield stress for other stress states are evaluated by interpolation
as with any finite element representation.  
The reference value of yield stress can be re-scaled to account for rate dependence.
The direction of plastic strain is readily determined from the normal to the yield surface.
The mesh can be refined in regions of higher curvature to better capture `corners' of the yield surface.
This requires that additional data be computed to provide values of the yield stress at the nodes of 
the refined mesh.


\section{Estimating macroscopic yield stress under multiaxial loading}
\label{sec:yield_detect_formulation}

As previously stated, the purpose of the methodology presented in this paper is to provide synthetic yield data for defining an anisotropic yield condition at the continuum, or macroscopic, scale taking into consideration the microstructure of the material.
Here, microstructure is meant to include, but not be limited to, the phase topology and contiguity, the grain morphology and  the crystallographic texture.  
Single crystal properties are assumed known for all of the structurally significant phases.  The crystal elastic moduli and the strengths of the potentially active slip systems are the primary requirements.  

The challenge for predicting macroscopic yield of a specimen is in determining at what load a plastic band extends across an entire cross-section of the sample.  In a polycrystalline sample, yielding commenses locally (within individual crystals) at various points in the polycrystal.  The extent of the yielding increases with increasing load, eventually reaching a point at which zones of plasticity interconnect to form a surface that completely cuts across the sample.  The load at which this occurs is taken here to be the macroscopic yield stress as there is no longer an elastic pathway connecting the load bearing ends of the sample.   Detecting the extinction of an elastic pathway itself is challenging, but can be 
tackled with more than one approach, as discussed later.   First, we focus on being able to predict the 
the macroscopic load at which this happens for a polyphase, polycrystalline sample.  

To determine an entire yield surface for an anisotropic material, values of the macroscopic yield stress are needed for an array of loading modes.  This array must span  the possible combinations of stress components that may be active.  In general, all five components of the deviatoric stress must be probed.  For special stress states, like biaxial loading under plane stress (in-plane loading of sheets), fewer than five may be sufficient.    A brute force approach would be to simulate the loading of a virtual sample under various combinations of the active stress components to a level sufficent to reach macroscopic yielding for each combination.
If one could instead predict the stress level corresponding to macroscopic yielding based on knowledge of the microstructural properties and a simple elastic simulation, the effort evaluate a yield surface could be substantially reduced.
The new methodology reported here addresses the latter approach, but also compares to the former.  

The new methodology builds on two developments: a multiaxial strength-to-stiffness parameter and a correlation between this multiaxial strength-to-stiffness parameter and the onset of local yielding. 
Using the strength-to-stiffness parameter together with the correlation for the prediction of local yielding, it is possible to predict the onset of yielding of finite elements of a virtual polycrystal  with only the elastic strain data from a single increment of a purely elastic finite element simulation plus knowledge of the single-crystal yield surface.  
These developments are reported in prior articles, \cite{pos_daw_multiaxial-y2e} and \cite{pos_daw_twophase}, in which  demonstrations are given for the yielding of a single-phase stainless steel (AL6XN) and  a duplex stainless steel (LDX-2101).
The new methodology entails the following steps:
\begin{enumerate}
    \item Using quantitative microscopy data, build a virtual sample with representative microstructure. This includes the phase volume fractions and topoology, the grain size distribution, grain shape distribution, and crystallographic orientations.  Assign appropriate properties to the grains, including the elastic moduli and strengths of the potentiallly active slip systems. 
    \item Establish a list of the loading conditions that span the section of stress space to be interrogated (full five dimensions of deviatoric stress space or a more limited subspace, such as biaxial stress states related to a  particular material-based reference frame).
    \item Perform simulations in which an increment of load is applied to the virtual sample.  The increment of load is sufficiently small that the sample remains elastic everywhere.  One simulation is performed for each of the loading cases identified in Step 2.
    \item  Using the data from the simulation, compute the strength-to-stiffness ratio for every finite element within every crystal of the mesh.
    \item  Using the yield prediction correlation, identify those finite elements that are likely to yield for a target estimate of the macroscopic yield stress.
    \item  Using yield detection algorithm, determine if macroscopic yielding has occurred at the target stress. 
    \item  Correct the target until it is the lowest macrocopic stress at which macroscopic yield is detected.
\end{enumerate}

In the remaining paragraphs of this section, the strenth-to-stiffness parameter, the yield prediction correlation, and the yield detection algorithm are briefly summarized.  Citations to articles with
more detail are provied.

\subsection{Strength-to-stiffness ratio for multiaxial stress states}\label{sec:Y2E_background}

A thorough development of the multiaxial strength-to-stiffness parameter is available in \cite{pos_daw_multiaxial-y2e}; here a brief summary is provided.  
  The strength-to-stiffness parameter, $r_{SE}$, draws the strength from the single crystal yield surface and the stiffness from Hooke's law.  
Equation~\ref{eq:rss_projection} defines yielding in single crystals.    This equation states that yielding occurs when the magnitude of the resolved shear stress on any slip system is equal to the critical resolved shear stress $\tau^\alpha_{cr}$ for that slip system.   
     The yield condition may be restated in an equivalent form as 
\begin{equation}\label{eqn:y2e_YieldCondition}
\max_\alpha \left( \frac{\vert \tau^\alpha \vert}{\tau^\alpha_{cr}} \right) = 1
\end{equation}
Let $(\cdot)^*$ denote a slip system where 
\begin{equation}\label{eqn:y2e_TauStar}
\frac{\vert \tau^* \vert}{\tau^*_{cr}} \equiv \max_\alpha \left( \frac{\vert \tau^\alpha \vert}{\tau^\alpha_{cr}} \right)
\end{equation}
Equation~\ref{eqn:y2e_TauStar} embodies both the strength ($\tau^*_{cr}$) and the stress ($\tau^* $) acting on the crystal.  The stiffness can be inferred from the stress if the elastic strain is known.  To construct the strength-to-stiffness from Equation~\ref{eqn:y2e_TauStar}, the stiffness is introduced by dividing the stress by an appropriate estimate of the elastic strain.  Here,  $E_\mathit{eff}$, defined as the effective macroscopic elastic strain is used for this purpose.  This gives
\begin{equation}\label{eqn:y2e_Y2EEval}
r_{SE} = E_\mathit{eff} \frac{\tau^*_{cr}}{\vert \tau^* \vert}
\end{equation}
Equation~\ref{eqn:y2e_Y2EEval} may be evaluated for a macroscopic stress in the elastic regime along the load path of interest. The crystal deviatoric stress needed to evaluate $\tau^*$ can be evaluated, for example, from one load increment of a purely elastic finite element simulation.  
$r_{SE}$ is a very simple, but quite general, multiaxial estimate of the strength-to-stiffness. 



\subsection{Methodology for predicting local yield}\label{sec:yield_prediction}

The multiaxial strength-to-stiffness parameter can be used to predict the macroscopic stress at which a region will yield. The same assumptions about the macroscopic and local load histories that were used for the strength-to-stiffness formulation are applied~\cite{pos_daw_multiaxial-y2e}. In this analysis, the incremental load is carried only by elastic regions as regions that have yielded as assumed to be unable to carry additional load (neglecting hardening over the elastic-plastic transition). Designating the local (crystal) stress as $\sigma$ and the macroscopic average stress as $\Sigma$, the ratio of local stress increment to macroscopic stress increment is $\Delta\sigma / \Delta\Sigma$  and is  zero for plastic regions. With the macroscopic stress increasing over the elastic-plastic transition, this ratio must increase for the remaining elastic regions as yielding progresses. As developed in \cite{pos_daw_multiaxial-y2e}, the local stress increment is approximated as being directly proportional to the applied macroscopic stress increment and inversely proportional to the elastic volume fraction raised to an empirical power $n$
\begin{equation}\label{eqn:local_stress_incr}
\Delta \sigma \varpropto \left( \frac{v}{v^e} \right)^n \Delta \Sigma
\end{equation}
where $v^e$ and $v$ are the elastic and total volumes, respectively. The volumetric scaling is an empirical factor that models the increase in local load increment, relative to the macroscopic load increment, for elastic regions that occurs as yielding propagates through an aggregate. The resolved shear stress is a projection of the local stress and therefore exhibits the same scaling.

\begin{figure}[ht]
\centering
\subfigure[$BR = 0.00$]{\includegraphics[trim = 0in 0in 0.9in 0in, clip]{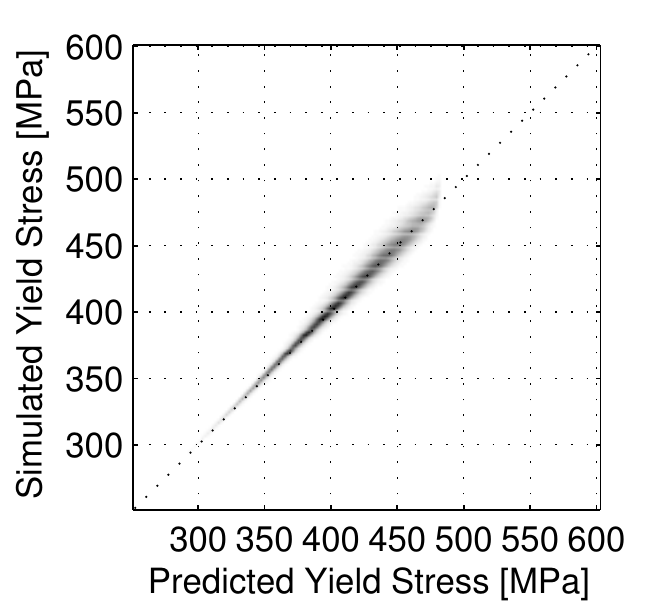}
\label{fig:ElemYldStress-BR000}} \qquad
\subfigure[$BR = 0.25$]{\includegraphics[trim = 0in 0in 0.9in 0in, clip]{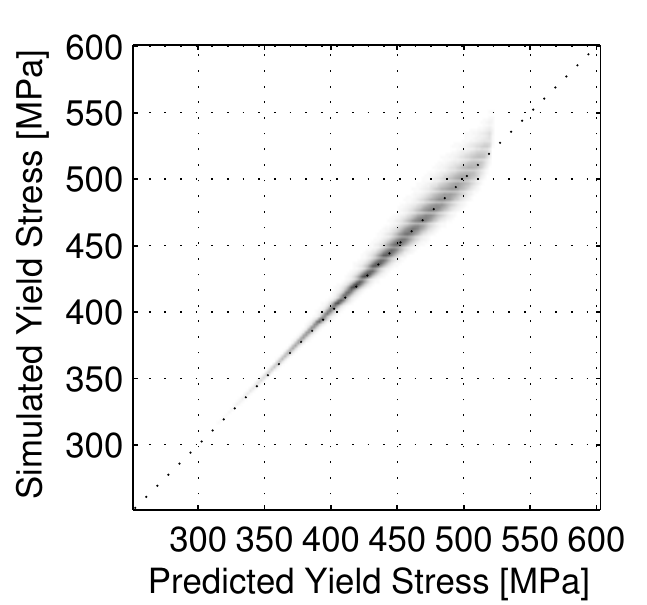}
\label{fig:ElemYldStress-BR025}}
\subfigure[$BR = 0.50$]{\includegraphics[trim = 0in 0in 0.9in 0in, clip]{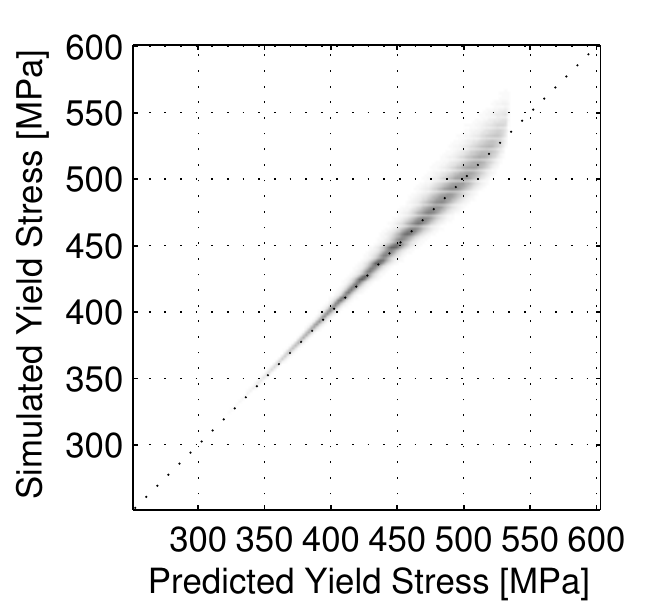}
\label{fig:ElemYldStress-BR050}} \qquad
\subfigure[$BR = 0.75$]{\includegraphics[trim = 0in 0in 0.9in 0in, clip]{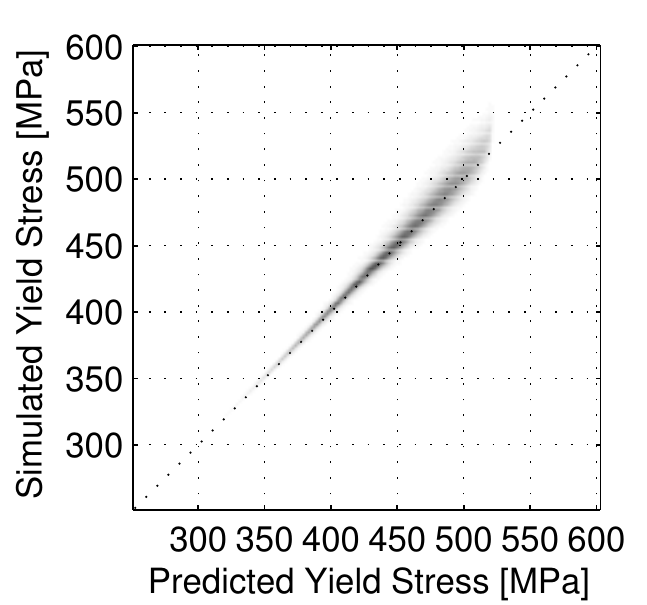}
\label{fig:ElemYldStress-BR075}}
\subfigure[$BR = 1.00$]{\includegraphics[trim = 0in 0in 0.9in 0in, clip]{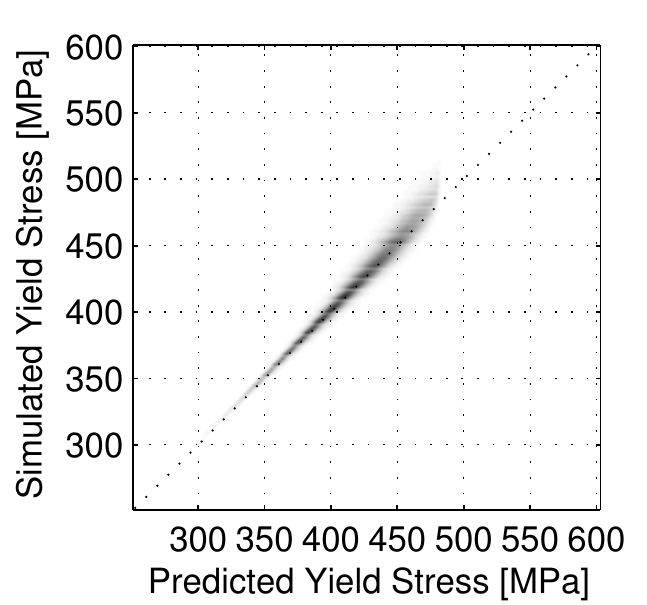}
\label{fig:ElemYldStress-BR100}} \qquad
\subfigure{\includegraphics[trim = 2.6in 0in 0in 0in, clip]{SimPredElemYldStress-BR000.pdf}}
\caption{Simulated and predicted elemental yield stresses.  Reproduced from \cite{pos_daw_twophase} }\label{fig:SimPredElemYldStress}
\end{figure}

Introducing the constant of proportionality $\left( d \tau^* / d \Sigma \right)_0$ yields the equality
\begin{equation}\label{eqn:drss_dsigbar}
\frac{\Delta \tau^*}{\Delta \Sigma} = \left( \frac{d \tau^*}{d \Sigma} \right)_0 \left( \frac{v}{v^e} \right)^n
\end{equation}
where $\left( d \tau^* / d \Sigma \right)_0$ is the derivative of the resolved shear stress with respect to the macroscopic stress coefficient at zero load. This derivative can be evaluated from one load increment of a finite element simulation in the completely elastic regime. Equation~\ref{eqn:drss_dsigbar} can be integrated numerically by summing over the number of load steps ($N$) to calculate the macroscopic stress at which a region yields, corresponding to $\tau^* = \tau_{cr}^*$ in the rate-independent limit
\begin{equation}\label{eqn:rss_integrate}
\tau^* = \sum_{i=1}^N \frac{\Delta \tau^*}{\Delta \Sigma} \Delta\Sigma_i
\end{equation}
When evaluating the macroscopic stress at which elements in a finite element mesh yield, the elements are first binned according to strength-to-stiffness ratio to reduce numerical errors due to summing many small numbers. Each bin is associated with a range of macroscopic stresses over which all elements in the bin undergo yield. This range defines the macroscopic stress increment $\Delta\Sigma_i$ for the bin. The average elastic volume is used for the numerical integration.
For the volumetric scaling exponent, $n$, a value  of $2/3$  was demonstrated to give an excellent fit between the predicted value for yield from Equation~\ref{eqn:rss_integrate} and the yield extracted from simulation for single-phase stainless steel~\cite{pos_daw_multiaxial-y2e}. 
The predictive capability of the correlation is shown in Figure~\ref{fig:SimPredElemYldStress} for the 
case of the duplex stainless steel (LDX-2101) for a number of biaxial loading conditions.  The biaxial ratio, $BR$, is defined as the ratio of the lateral stress component to the axial stress component, ranging from zero for uniaxial tension to unity for balanced biaxial tension.
The plots show the correlation between the values of macroscopic yield provided by the method enumerated above and the values obtained by running simulations to loads sufficient to induce macroscopic yielding.

\subsection{Detecting macroscopic yielding}\label{sec:macro_yield_detection}

A sample under load is taken to have reached macroscopic yielding when a sufficient combination of regions of local plasticity interconnect to form a surface that spans any load bearing crossection of the sample.   The crossection need not be planar nor have any particular orientation relative to sample axes.
An equivalent manner to define the onset of macroscopic yielding is to require that no elastic ligament can be found that spans the sample between the load bearing surfaces.  Satisfying these 
requirements implies the existence of a yield band within the sample.    Detecting the load at
which a yield band first forms necessitates examining the spatial arrangement of the regions
of local plasticity.  Two such methods were considered, the existing method known as a flood fill algorithm~\cite{Shane_2016} and a new method based on eigenmodes of an elastic surrogate.  

A flood-fill algorithm is a geometric-based method that can be used to test for the presence of a yield band. The flood-fill algorithm determines whether two opposing mesh surfaces are connected via the
solid or structure within the boundary of the body. 
Here, the body between the surfaces is comprised of elements of the mesh that have yet to yield locally; elements with active plasticity are ignored.  
Two disconnected surfaces implies the existence of a yield band because there is no longer
an elastic ligament connecting the surfaces. 
In using the flood-fill algorithm here, three different criterion for element connectivity were considered: elements share at least one face, elements share at least one edge, and elements share at least one vertex. Of these three connectivity criterion, the vertex connectivity is most analogous to the eigenmode formulation, discussed next, and was therefore used in the analysis.

The second method of macroscopic yield detection is based on the eigenmodes of an elastic surrogate. 
As with the flood fill agorithm, any element with active plasticity is excluded from the computation.
The finite element model of the elastic surrogate will have eigenmodes and associated eigenvalues that characterize its structural behavior. In the event that there are eigenmodes with zero-valued eigenvalues, a singularity exists that imples a zero-work mode of deformation is possible.  
If such a mode corresponds to separation of the loading surfaces, then a yield band has formed. 
This approach provides a clear numerical test, but the challenge in applying the aprroach is 
to separate global zero-work modes from local zero-work modes.
More detail is provided in the appendix. 

Either detection algorithm can be used with either the simulated or predicted distributions of yielding, described in Section~\ref{sec:yield_surf}.  The two methods were found to give comparable estimates of the onset of macroscopic yielding.    For the application presented here the flood fill agorithm is used as it is more computationally efficient. 



\clearpage

\section{Finite element model for LDX-2101 stainless steel }
\label{sec:simulation_virtual-px}
To carry out the new methodology delineated in Section~\ref{sec:yield_detect_formulation}, 
a virtual sample of the subject material must be instantiated and its response to various
loading states simulated.   The simulations are conducted with finite element analyses, here 
done using the code, \fepx.
\fepx\, has the capability to model the elastoplastic deformations of a polycrystalline aggregate to large strains using anisotropic elasticity and restricted slip plasticity for the constitutive models.
The simulations need only involve a single loading increment within the elastic domain, however, to follow the new method's protocol. Loading into the regime of fully-develop plastic flow was carried out, nevertheless, so that comparisons of macroscropic yield stress could be made between values using the new methodology  and values determined by a suite of simulations taken into the fully plastic regime.

In this section the finite element simulations of LDX-2101 are discussed.  
First, the basic capabilities of \fepx\, are listed.  To conduct  simulations using \fepx, it is first necessary to build a virtual sample and assign attributes to it.  The attributes include the mechanical properties of each phase plus values for the initial state, namely the strengths of the phases and the orientations of the crystal lattices of the grains.   This section documents the choices made in constructing a suitable virtual sample for LDX-2101.  In reaching this sample, a comprehensive  study was carried out previously to examine the sensitivities of the lattice strains to the sample instantiation and the single crystal elastic and plastic constitutive parameters.  This study is reported in a separate paper~\cite{pos_daw_parmstudy}.   The study showed that the macroscopic stress-strain and fiber-averaged lattice strain responses for LDX-2101 were relatively insensitive to microstructure and justified the use of the simplified mesh to extract these particular quantities of interest.  Further, it demonstrated that the single crystal constitutive parameters were capable of predicting the mechanical response well for the uniaxial loading case.   Finally, the list of loading conditions is provided.  The loading conditions span a range of biaxiality possible under plane stress.  The procedure for loading samples under biaxial stress using \fepx\, is discussed.

\subsection{Finite element framework}\label{sec:fe_framework}

\fepx\, was developed for modeling the response of virtual microstructures over large strain motions and incorporates the following capabilites:
\begin{itemize}
\item{nonlinear kinematics capable of handling motions with both large strains and large rotations;}
\item{anisotropic elasticity based on cubic or hexagonal crystal symmetry;}
\item{anisotropic plasticity based on rate-dependent slip on a restricted number of systems for cubic or hexagonal symmetry;} 
\item{evolution of state variables for crystal lattice orientation and slip system strengths.} 
\end{itemize}
To accommodate these behaviors the finite element formulation has incorporated a number a numerical features, such as:
\begin{itemize}
\item{higher-order, isoparametric elements with quadrature for integrating over the volume;} 
\item{implicit update of the stress in integrations over time;}
\item{monotonic and cyclic loading under quasi-static conditions.}
\end{itemize}

The virtual microstructures are instantiated to define grains and to discretize the grains into multiple 10-node tetrahedral finite elements. 
The local behaviors associated with the material within an element correspond to those of a single crystal. 
A phase and initial crystallographic orientation are assigned to each element. 
Each element's state, characterized by its elastic strain, crystallographic orientation, and slip system hardnesses, evolves independently from that of the other elements. 
Velocity, traction, or mixed boundary conditions are applied to each surface of the virtual specimen. 
Compatibility of the deformation is enforced by the smoothness of motion guaranteed by the continuity of the interpolation functions. 
The deformation satisfies quasi-static equilibrium, which is enforced via the weak form of conservation of linear momentum. 

An implicit time integration scheme is used to solve for the velocity field and material state.  An estimated material state at the end of each time increment is used to evaluate the constitutive equations. Implicit integration ensures stability. The solution algorithm for each time increment begins with initializing an initial guess of the velocity field. The deformed geometry at the end of the time increment is then estimated, based on the velocity field. The velocity gradient is computed and used to solve for the crystal state at each quadrature point.  Iteration continues on the geometry, crystal state, and velocity field until the velocity solution is converged. The solution then advances to the next time increment.

\subsection{Virtual microstructure instantiation}\label{sec:instantiation}

LDX-2101 microstructure presents several challenges for mesh generation. The first challenge is the irregular shapes of the ferrite grains. Austenite grains transform from the ferrite matrix along grain boundaries. This leaves behind irregularly-shaped ferrite grains. 
To address these challenges, idealized microstructures were created using hexagonal prismatic building blocks. 
Detail of the microstructure are discussed in greater detail in \cite{pos_daw_twophase}.

Microstructure generation mimics the material processing.
Meshes were constructed by tessellating a hexagonal prismatic base mesh, comprised of multiple tetrahedral elements. The hexagonal prismatic base regions formed building blocks that were then stitched together to form grains. Annealing twins are just on the order of mesh resolution and were not represented in the virtual microstructure.  
Following the generation of the underlying mesh, a two-dimensional microstructure is created. The two-dimensional parent ferrite microstructure is generated using mapped Voronoi tessellation. Austenite regions are then randomly transformed out along grain boundaries, until the desired volume ratio between the two phases is obtained. The two-dimensional microstructure is then extruded to form a three-dimensional microstructure. Each resulting columnar region is divided into grains based on a grain size distribution. All the grains in a column are assigned to the same phase as that of the parent two-dimensional region. Grain orientations are assigned by randomly sampling the experimentally-determined orientation distribution function for each phase.
See~\cite{pos_daw_parmstudy} for additional details regarding the sensitivity of the 
results to the sample instantiation.

\begin{figure}[ht]
	\centering
	\subfigure[Grain]{\includegraphics[width=0.49\linewidth]{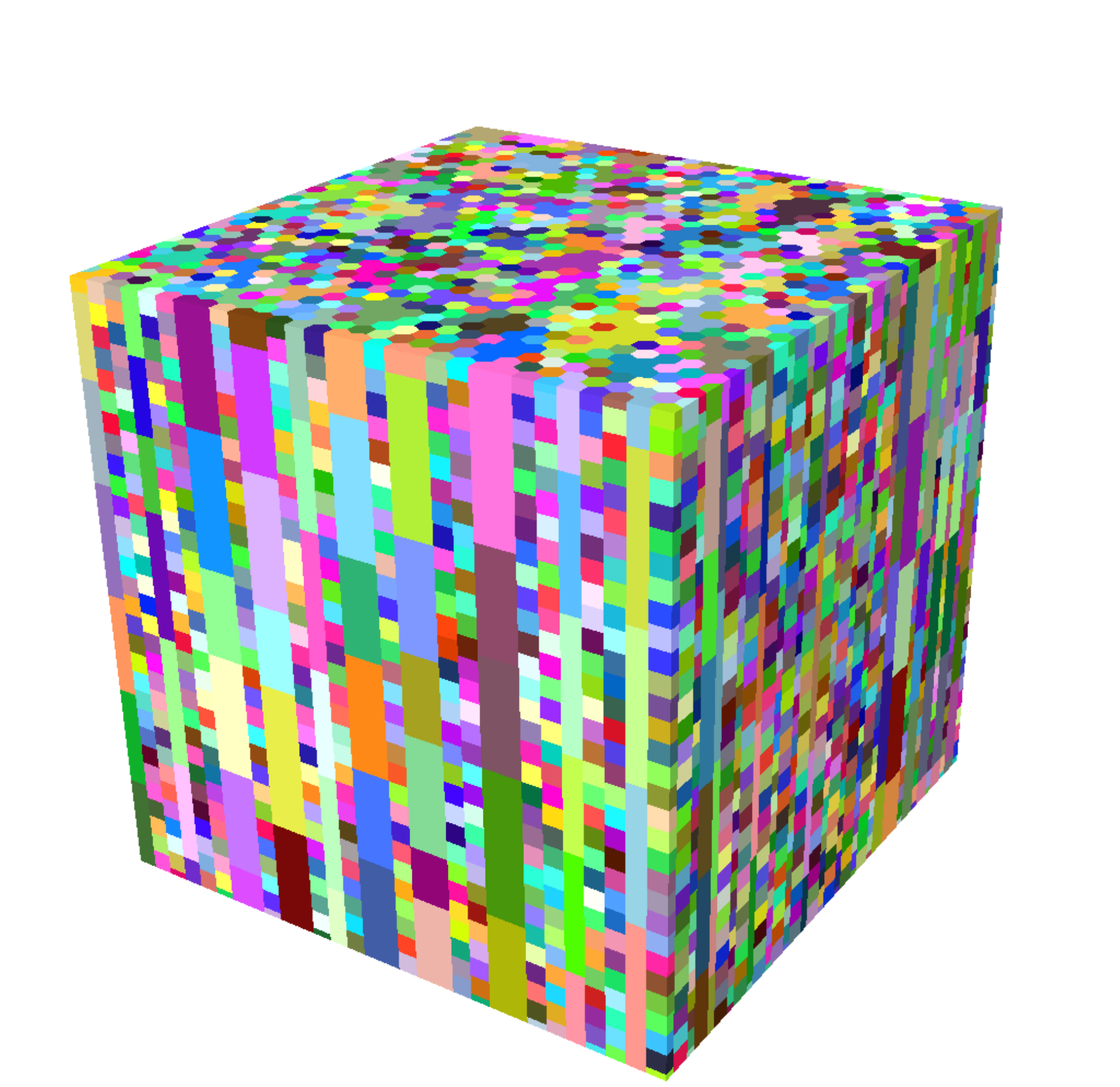}\label{fig:hex30-90-1-2-grain.pdf}}
	\subfigure[Phase]{\includegraphics[width=0.49\linewidth]{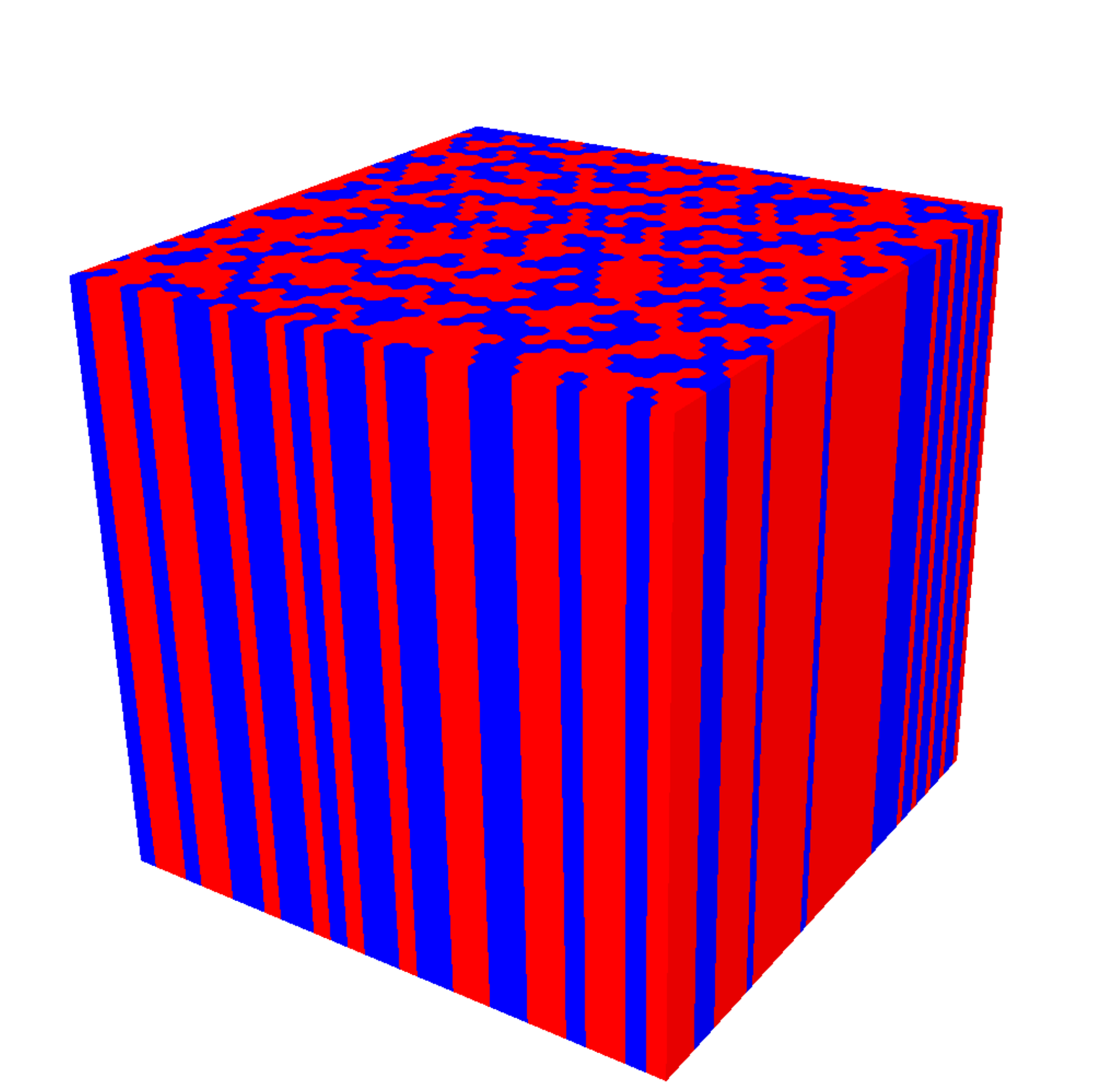}\label{fig:hex30-90-1-2-phase.pdf}}
	\caption{Representative virtual microstructure for LDX-2101, exhibiting a columnar phase structure with elongated, irregularly-shaped ferrite grains.}
	\label{fig:StandardMesh}
\end{figure}

This method of mesh generation effectively addressed the issue of meshing concave regions. It also ensures good mesh quality. However, it lacks adaptive mesh refinement around small features. The entire domain must be meshed with the same level of refinement as the smallest feature that is to be resolved. Large ferrite grains are meshed with the same refinement as small austenite grains, which makes the simulation more computationally intensive. Increased computation time is the tradeoff for having a representative microstructure with good element quality. A representative mesh consisting of 1,701,000 elements, 2,321,325 nodes, 18,673 austenite grains, and 769 ferrite grains is shown in Figure~\ref{fig:StandardMesh}. The level of mesh refinement for the smaller austenitic grains is sufficient to resolve intragranular stress and orientation gradients and is comparable to the level of mesh refinement in prior studies~\cite{Han05a, Wong10a, Marin12a}.

\clearpage

\subsection{Assignment of material parameters}\label{sec:parameters}

The constitutive model employed in \fepx\, is provided in the appendix.  This model requires elastic moduli as well as  parameters associated with the equations for restricted slip plasticity.  The values of these moduli parameters  were derived from multiple references and experiments. A comprehensive study of influence of the parameter choices on the simulated mechanical response was carried out previously for LDX-2101, as detailed in \cite{pos_daw_parmstudy}.

The single crystal elastic moduli were taken from the literature and are listed in Table~\ref{tab:ElasticConstants}. Ledbetter's elastic constants for austenitic stainless steel~\cite{Ledbetter01a} were used for austenite and Simmons and Wang's constants for pure BCC iron were used for ferrite~\cite{Simmons71a}.  These choices provided good matches to the lattice strains measured by neutron diffraction for a set of six crystallographic reflections (three for each phase). 
\begin{table}[ht]
	\centering
	\caption{Single crystal elastic constants using the strength of materials convention ($\tau_{44} = c_{44} \gamma_{44}$).}
	\begin{tabular} {c c c c}		
		phase & $c_{11}$ & $c_{12}$ & $c_{44}$ \\
		& (GPa) & (GPa) & (GPa)  \\ \hline
		FCC & 205 & 138 & 126 \\ 
		BCC & 237 & 141 & 116
	\end{tabular}	
	\label{tab:ElasticConstants}
\end{table}

Plasticity parameters were evaluated from a combination of physical experiments and simulations.  
Uniaxial strain rate jump tests were employed to characterize the rate sensitivity of plastic slip. 
Values of $m$ of 0.020 and 0.013 were found to provide good agreement with the measurements for the
austenite and ferrite, respectively.
Macroscopic stress-strain data for monotonic uniaxial loading at constant strain rate were used to characterize the remaining plasticity parameters.
In the parametric study \cite{pos_daw_parmstudy}, four combinations of initial strength and hardening rate were evaluated, the four combinations given by equal or unequal initial strength and equal or unequal hardening rate.  
For either the unequal initial strength or the unequal hardening rate combination, the ferrite value was twice the austenite value. 
Further, the specific values were chosen so that the volume-averaged values were the same for the 
equal and unequal cases.  
These combinations were motivated by prior research reported in the literature that examined the relative properties of the austenitic and ferrritic phases in stainless steels.  
The parametric study showed that the combination of equal initial strength and inital hardening rate overall matched the  lattice strains measured under uniaxial tension in which special attention was given to the 
progression of yielding as indicated by the inflection points in the lattice strain histories.
Consequently, in constructing the macroscopic yield surface here, equal initial strengths and hardening rates are specified. 
However, as there were some attributes of the comparisons for which an assumption of stronger ferrite provided the better comparison, a second yield surface is constructed for comparison.  For this case, the ferrite phase initial slip system strength is twice that of the austenite while the hardening rates are the same.  The rate sensitivities are the same for both cases.  
The two sets of plasticity parameters are listed in Table~\ref{tab:PlasticityParams}.

\begin{table}[ht]
	\centering
	\caption{Plasticity parameters for two cases: equal and unequal initial slip system strengths for the ferritic and austenitic phases. }	
	\begin{tabular} {c c c c c c c c}
	Case &	phase & $m$ & $n^\prime$ & $\dot{\gamma}_0$ & $h_0$ & $g_0$ & $g_s$ \\
	&	& & & ($\mathrm{s^{-1}})$ & (MPa) & (MPa) & (MPa) \\ \hline
	Equal Strength &	FCC & 0.020 & 1 & $10^{-4}$ & 336 & 192 & 458 \\ 
	&	BCC & 0.013 & 1 & $10^{-4}$ & 336 & 192 & 458 \\  \hline
	
	Unequal Strength&	FCC & 0.020 & 1 & $10^{-4}$ & 336 & 134 & 400 \\ 
	&	BCC & 0.013 & 1 & $10^{-4}$ & 336 & 269 & 535 \\  \hline
	\end{tabular}
	\label{tab:PlasticityParams}
\end{table}

\subsection{Loading modes simulated}\label{sec:triaxial}

A condition of plane stress often exists in thin sections, such as sheet metal during manufacturing processes  and the webs in semimonocoque structures.
Thin-walled tubes that are subjected to axial loads and internal pressure provide additional examples that arise both for load-bearing components and for samples used  in mechanical testing.  
In this latter example, various states of biaxiality
exist for different combinations of load and pressure.
  The stress tensor for these stress states can be written in the principal bases as:
\begin{equation}
\left[ \cauchy \right]
= \left[  \begin{array}{ccc}  \sigma_{xx} & 0 & 0 \\ 0 &  \sigma_{yy}  (=0)  & 0 \\ 0 & 0 & \sigma_{zz} \end{array} \right]
 \label{eqn:biaxial_stress_state}
\end{equation}
By adjusting the boundary velocities as described above, two components of the stress, $\sigma_{xx}$ and $\sigma_{zz}$, are controlled to give prescribed levels of stress biaxiality. 
The remaining component of the stress, $\sigma_{yy}$, 
is held fixed at zero. 
The biaxial ratio, $BR$, was introduced in Section~\ref{sec:yield_detect_formulation}.  It is defined as the ratio of the lateral stress component to the axial stress component, and ranges between zero for uniaxial tension and unity for balanced biaxial tension. 
Using the coordinate system in Equation~\ref{eqn:biaxial_stress_state}, the biaxial ratio is: $\frac{\sigma_{xx}}{\sigma_{zz}}$.
The prescribed ratios were chosen to give particular biaxial stress states ranging from simple uniaxial stress to balanced biaxial stress. 
Full simulations (taken through the elastoplastic transition) were conducted for five biaxial ratios: 0.0, 0.25, 0.5, 0.75 and 1.0.  
Additional simulations using the new predictive methodololgy only included biaxial ratios
of: $ \pm4{\rm,} \pm2 \,{\rm and} \pm1.33$.  
Also, the cases of $\pm\sigma_{xx}$ with $\sigma_{zz}=0$ were examined.

The simulation boundary conditions were defined to effect different levels of biaxiality over radial loading paths.
Symmetry boundary conditions were applied to the three orthogonal surfaces described by $x=0$, $y=0$, and $z=0$. Symmetry boundary conditions consist of zero velocity component in the the direction normal to the surface and zero traction component tangent to the surface. 
Mixed velocity/traction boundary conditions were applied to the other three orthogonal surfaces. On each surface, the normal component of velocity was prescribed uniformly over the surface, and the tangenetal component of traction was set to zero. The normal velocity on the $z$-surface was constant throughout the simulation to produce a constant engineering strain rate of $10^{-4} \,{\rm s}^{-1}$ in the axial $z$-direction. Iterations were performed on the normal components of velocity prescribed on the other two surfaces at each step of the simulation, so as to maintain the prescribed ratios of macroscopic stress components.

\clearpage

\section{Application to LDX-2101 stainless steel}
\label{sec:app2LDX}
The yield prediction method developed in Section~\ref{sec:yield_detect_formulation} was combined with the yield band detection algorithm to evaluate the biaxial yield surface for LDX-2101. Each point of the yield surface 
is an independent macroscopic stress state corresponding to a particular biaxial ratio.  Here, the yield surface
is represented with piecewise functions using interpolation functions and nodal point values. 
For each of the nodal point stress states, an elastic finite element simulation was conducted corresponding to 
the associated biaxial ratio. 
Following the procedure outlined in Section~\ref{sec:yield_detect_formulation}, the macroscopic stress at which each element yields was then calculated using the yield prediction method. 
This involves specifying a target value for the macroscopic stress coefficient, which served as a threshold to determine which elements were elastic and which had yielded. The yield band detection algorithm was then applied to determine whether the aggregate was yielded or elastic. The macroscopic stress coefficient was iteratively adjusted to home in on the macroscopic yield stress. This process was repeated for each nodal point of the finite element representation.
Typical results are first shown for the detection of a yield band under uniaxial stress loading.  This is followed by presentation of the entire yield surface for biaxial stress conditions.  Finally, the yield surface that results from assuming unequal phase strengths, also generated using the new methodology, is provided for comparison to the equal phase strength surface.

\subsection{Yield bands and their detection}
\label{sec:yield_bands}
Anisotropic properties at the grain level together with contrasting properties of the phases are the source
of spatial heterogeneity of deformation is polycrystalline samples subjected to loading, even with the loading
states are relatively simple (such as uniaxial or biaxial tension). Often referred to as grain interactions, 
this spatial attribute of the mechanical behavior is why microstructure must be considered in evaluating the
yield surface.  How the yielding in manifest over a sample depends on its microstructure and ultimately 
affects its yield surface.   For this reason, a yield band is first considered together with plots of 
the plastic deformation rate field to illustrate the heterogeneity of deformation expected in this alloy.

\begin{figure}[ht]
\centering
\subfigure{\includegraphics[width = 0.6\textwidth]{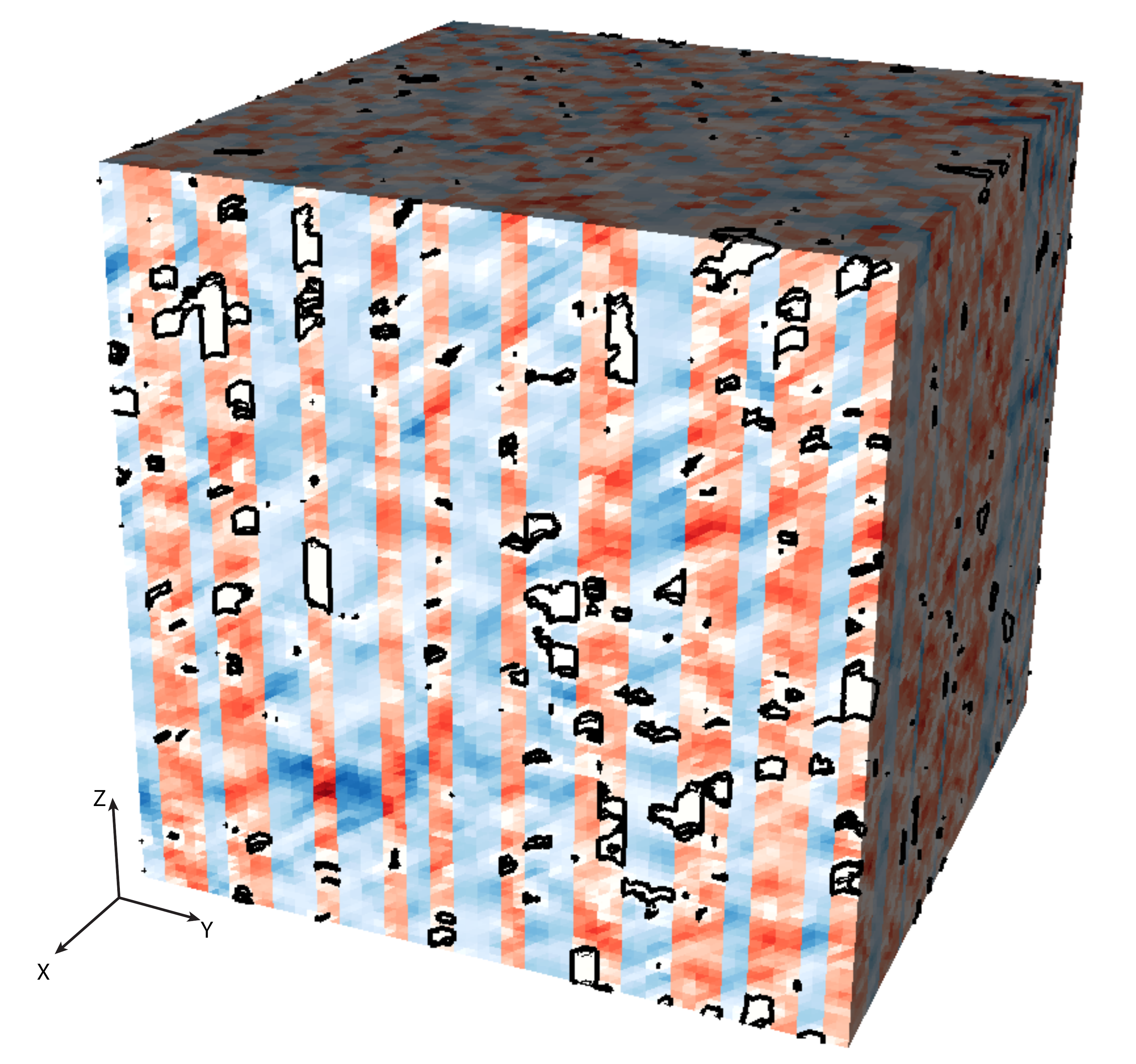}}
\subfigure{\includegraphics{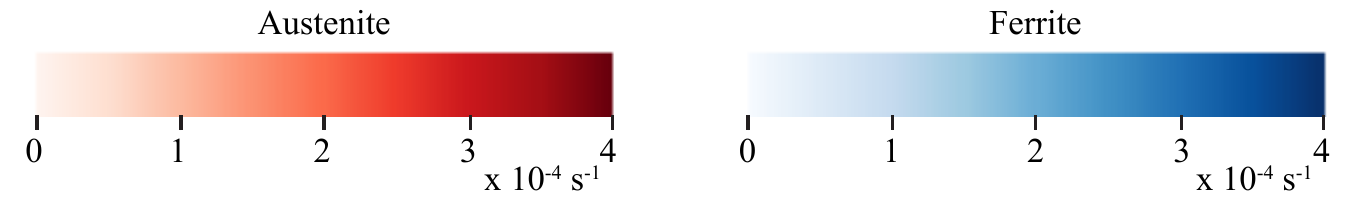}}
\caption{Effective plastic deformation rate at macroscopic yield for uniaxial loading. Contours corresponding to $D^p_{\mathit{eff}} = 10^{-5}~\mathrm{s}^{-1}$ are plotted. This threshold value corresponds to one-tenth of the macroscopic deformation rate.}
\label{fig:DPeff3D}
\end{figure}

\begin{figure}[ht]
\centering
\subfigure[$x=0.000$]{\includegraphics[trim = 2.5in 1.5in 0.9in 0.5in, clip, width = 0.30\textwidth]{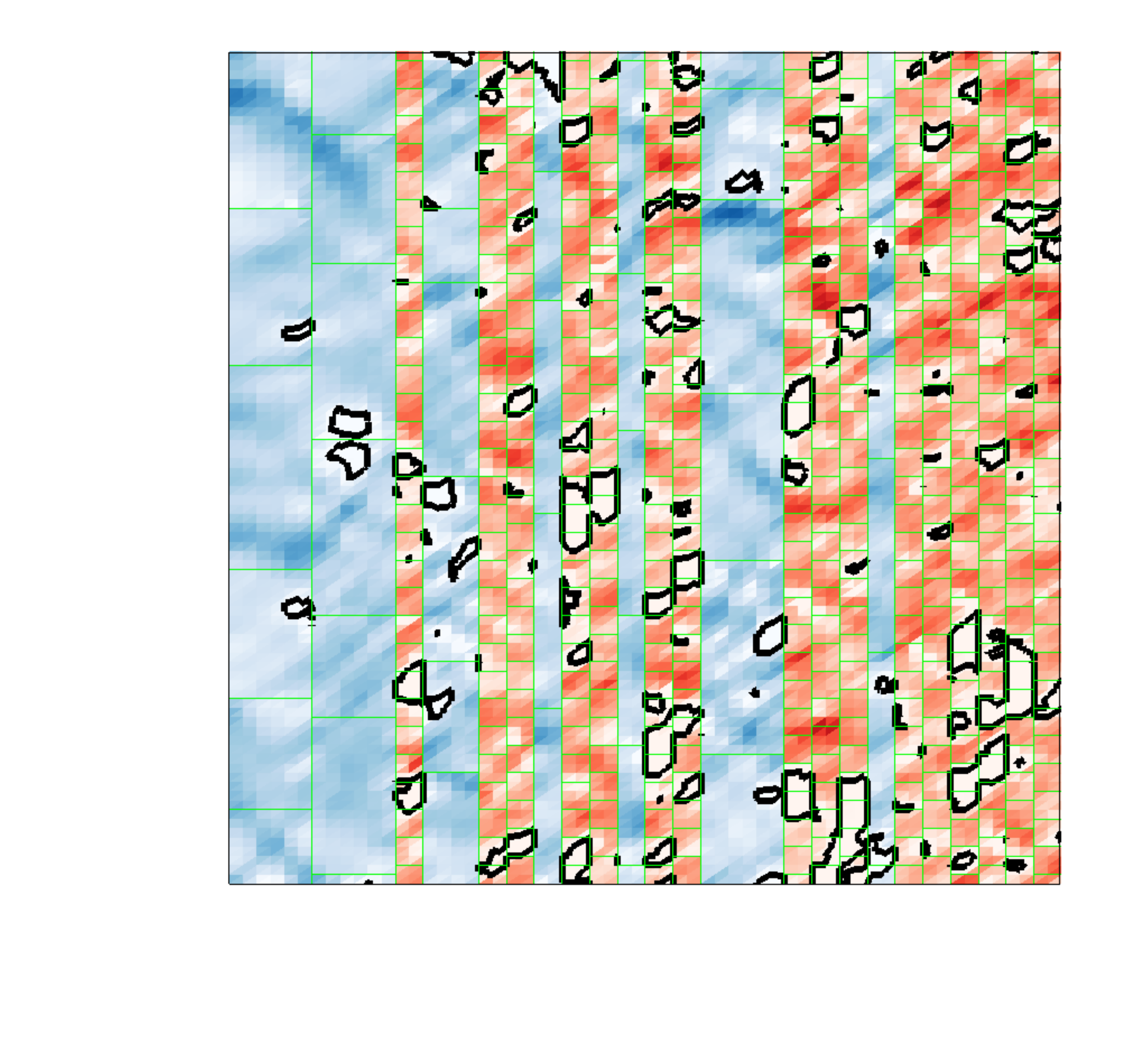}}
\subfigure[$x=0.125$]{\includegraphics[trim = 2.5in 1.5in 0.9in 0.5in, clip, width = 0.30\textwidth]{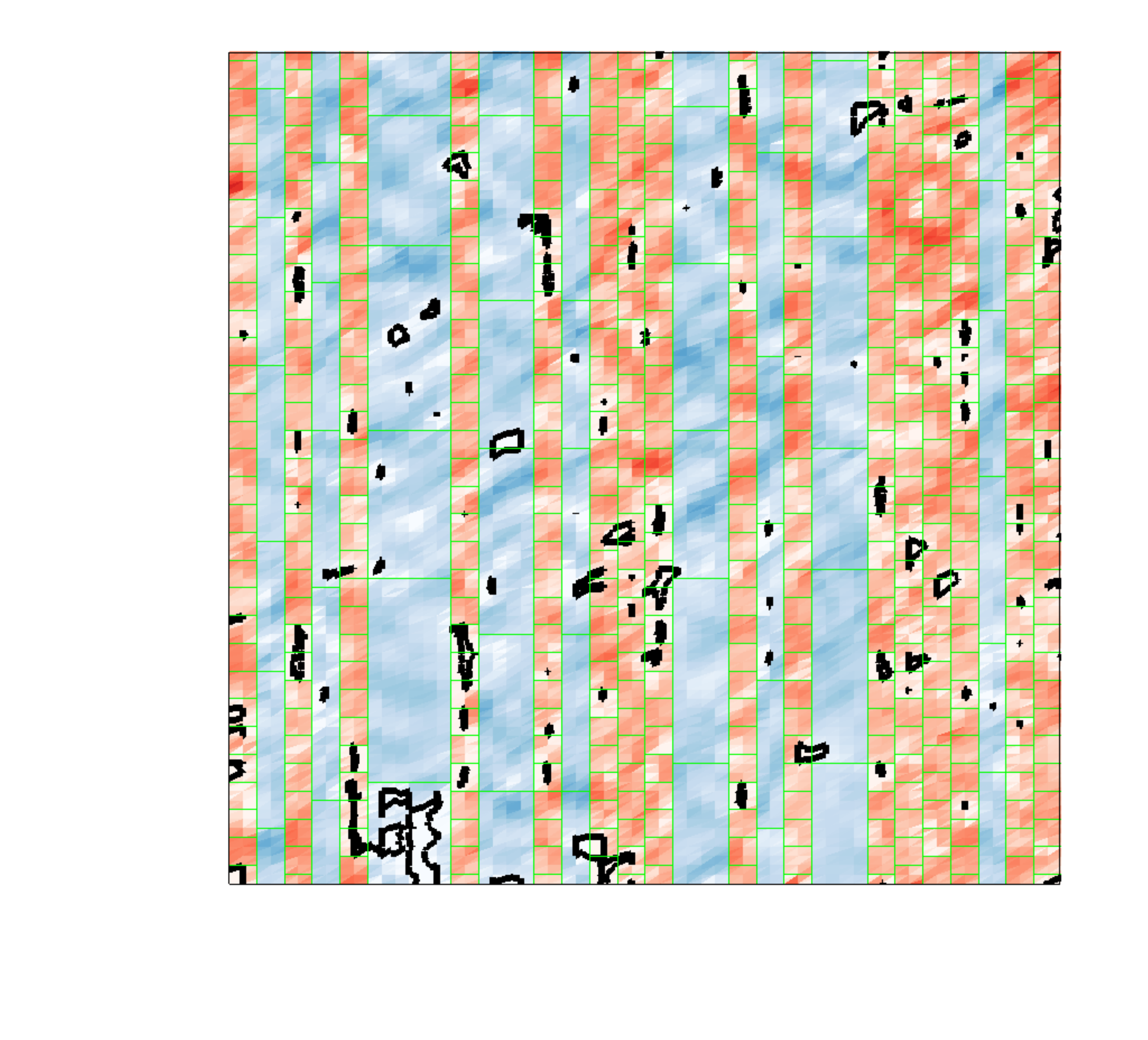}}
\subfigure[$x=0.250$]{\includegraphics[trim = 2.5in 1.5in 0.9in 0.5in, clip, width = 0.30\textwidth]{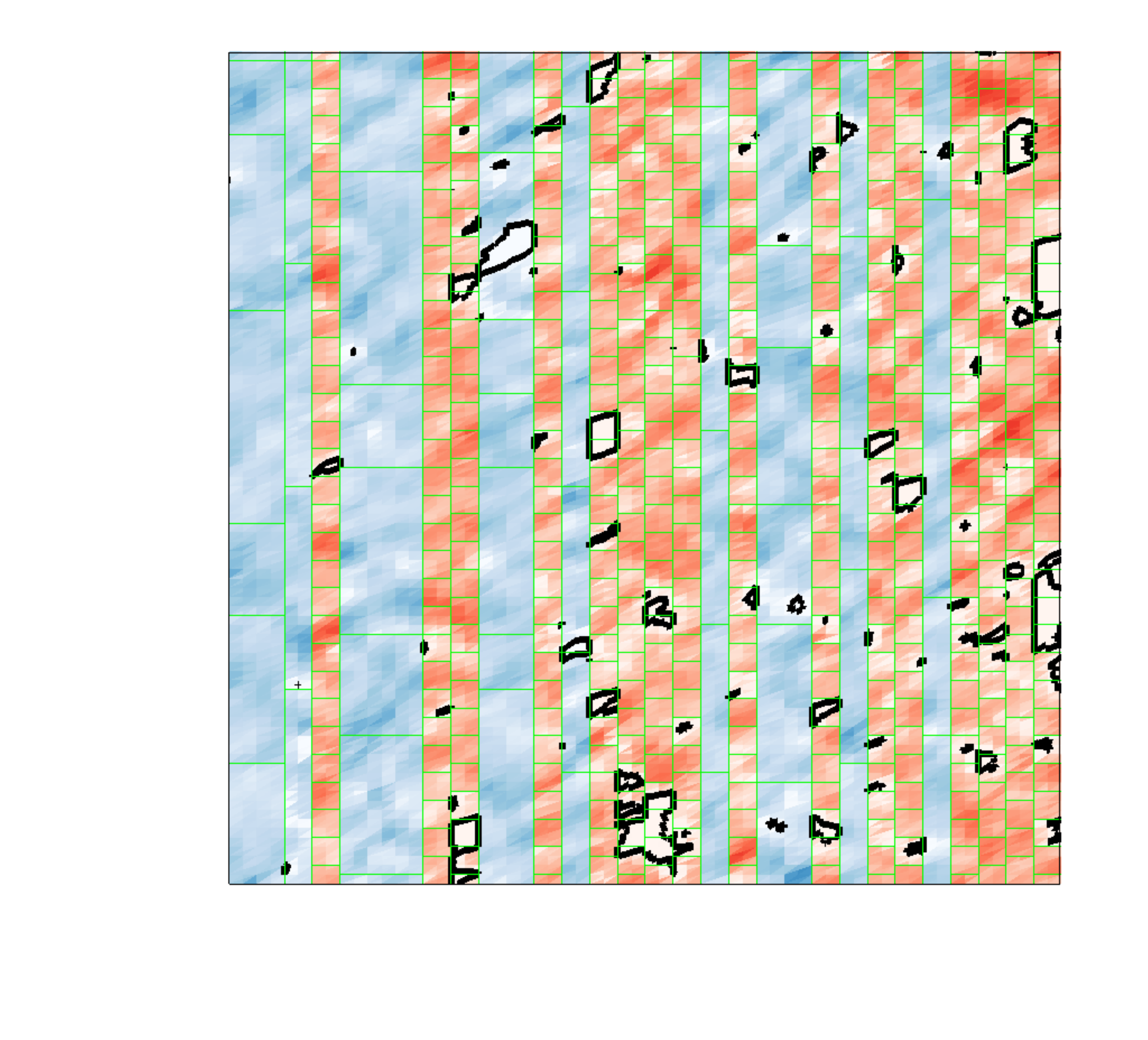}}
\subfigure[$x=0.375$]{\includegraphics[trim = 2.5in 1.5in 0.9in 0.5in, clip, width = 0.30\textwidth]{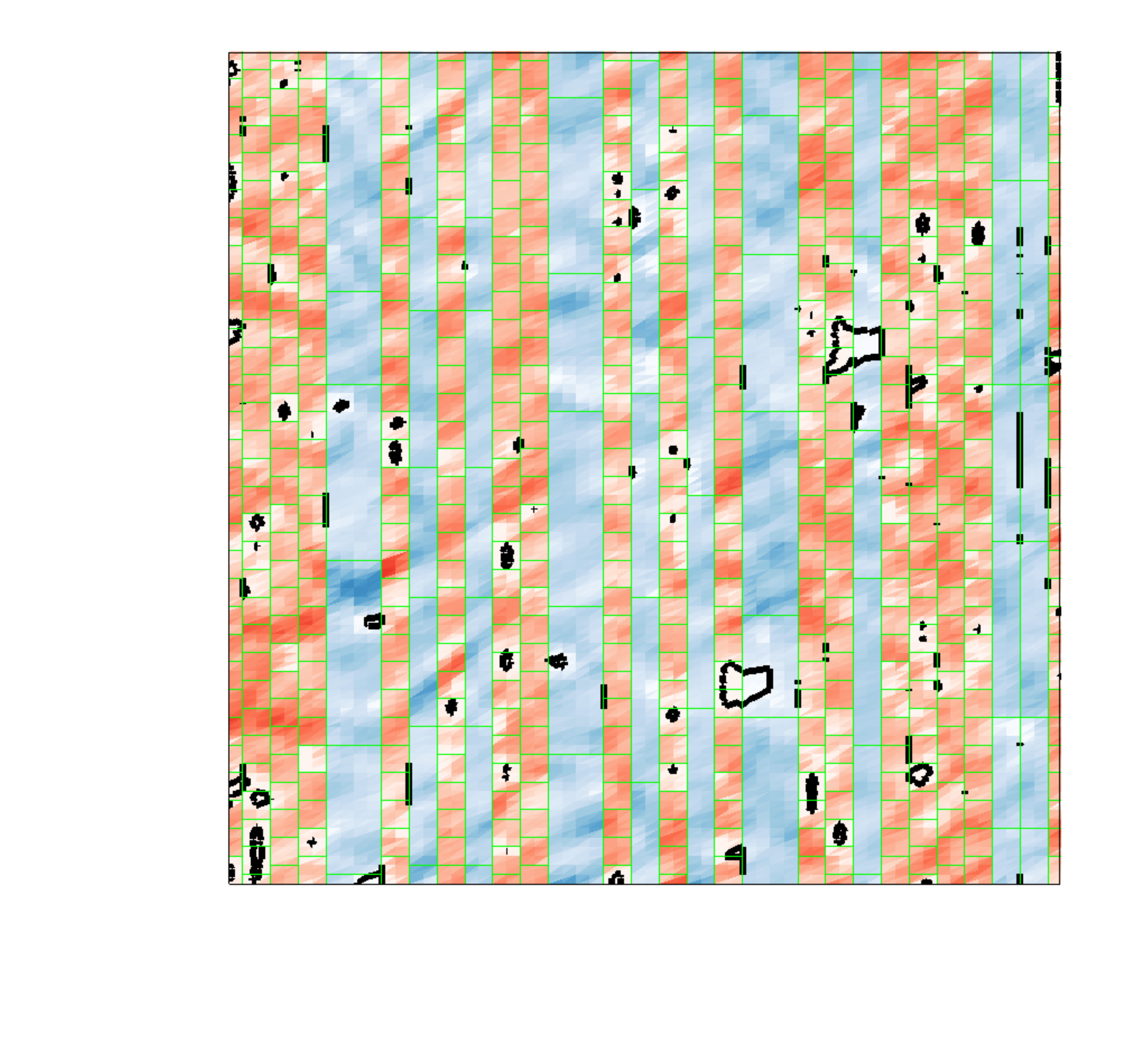}}
\subfigure[$x=0.500$]{\includegraphics[trim = 2.5in 1.5in 0.9in 0.5in, clip, width = 0.30\textwidth]{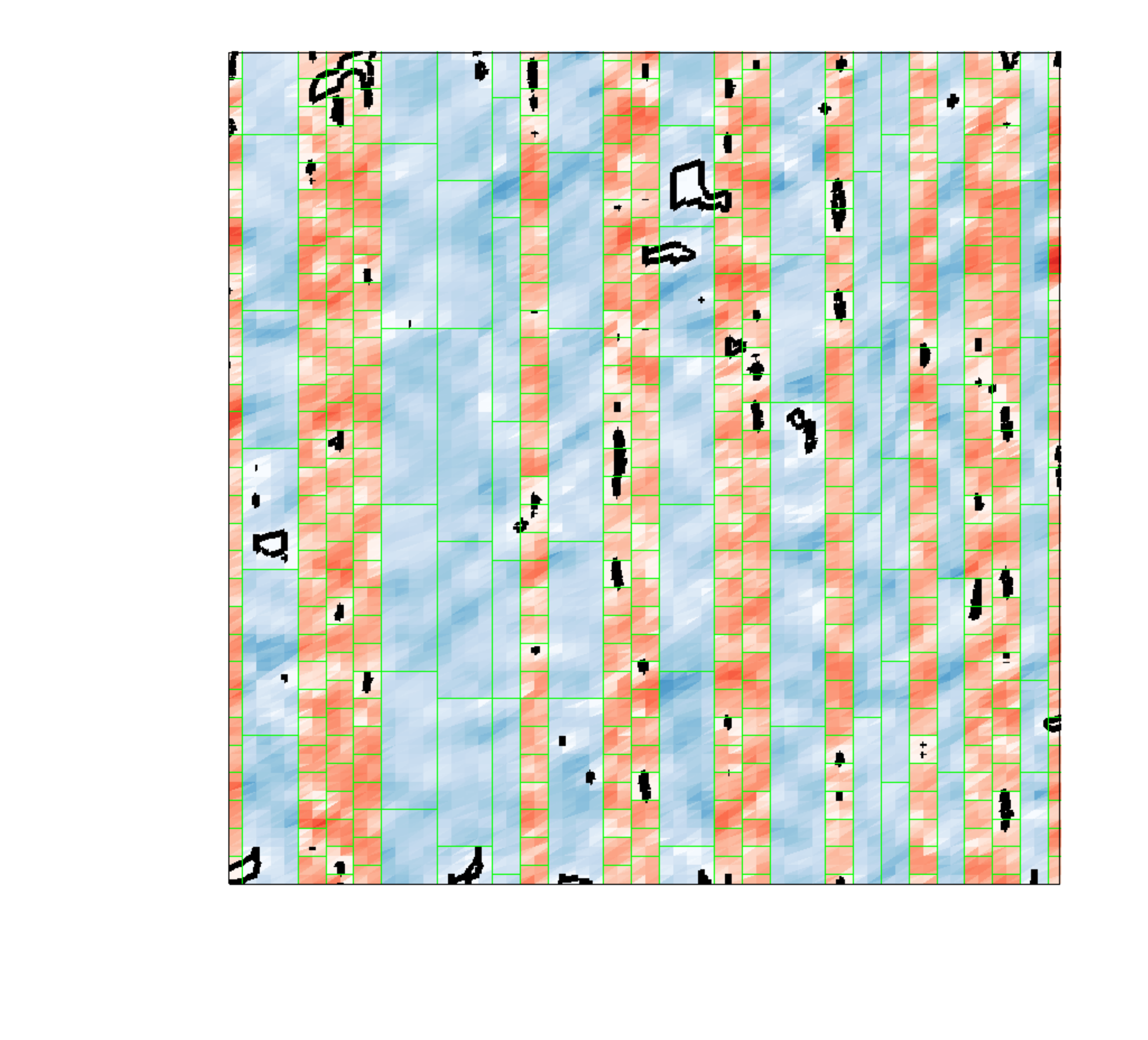}}
\subfigure[$x=0.625$]{\includegraphics[trim = 2.5in 1.5in 0.9in 0.5in, clip, width = 0.30\textwidth]{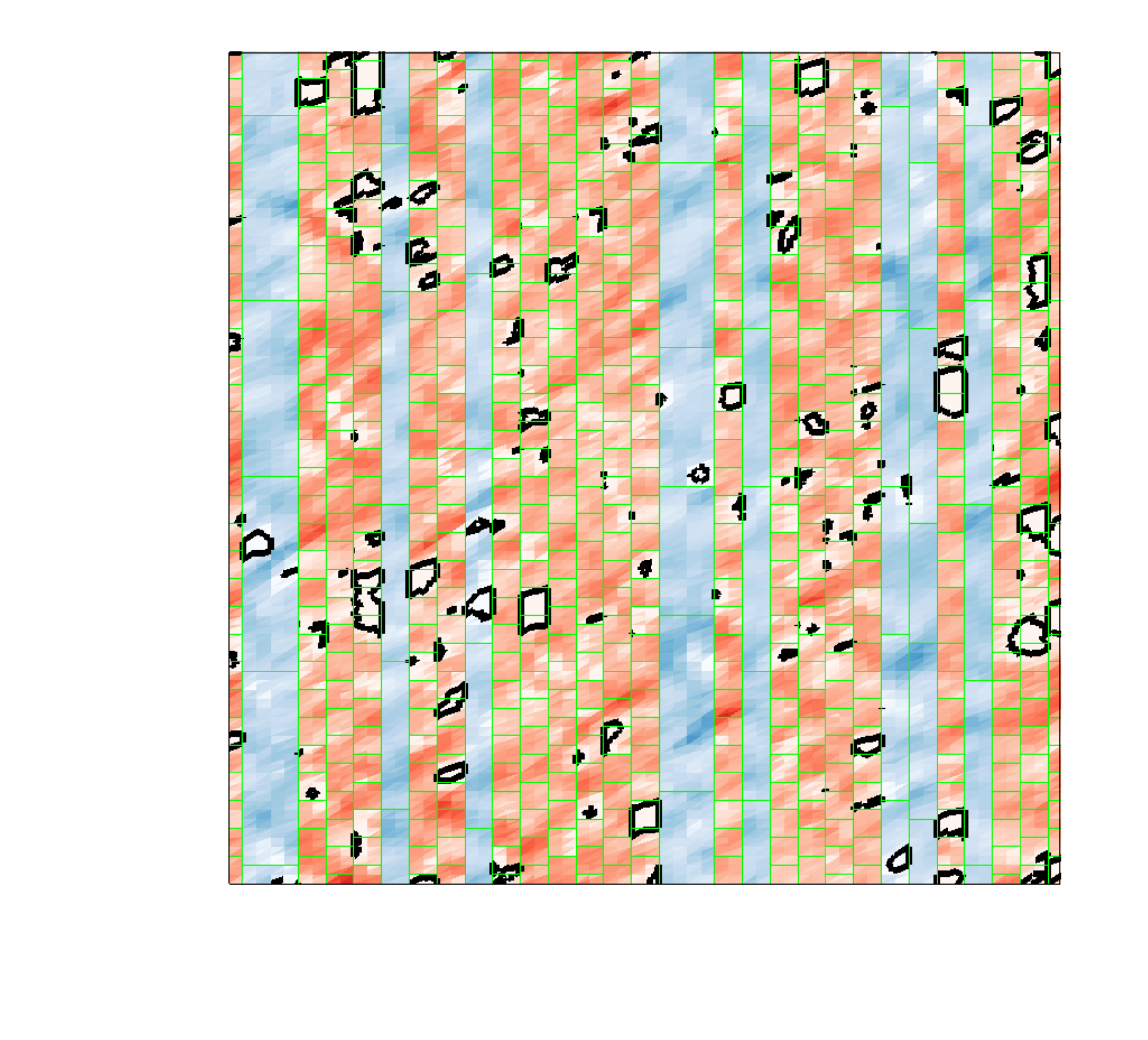}}
\subfigure[$x=0.750$]{\includegraphics[trim = 2.5in 1.5in 0.9in 0.5in, clip, width = 0.30\textwidth]{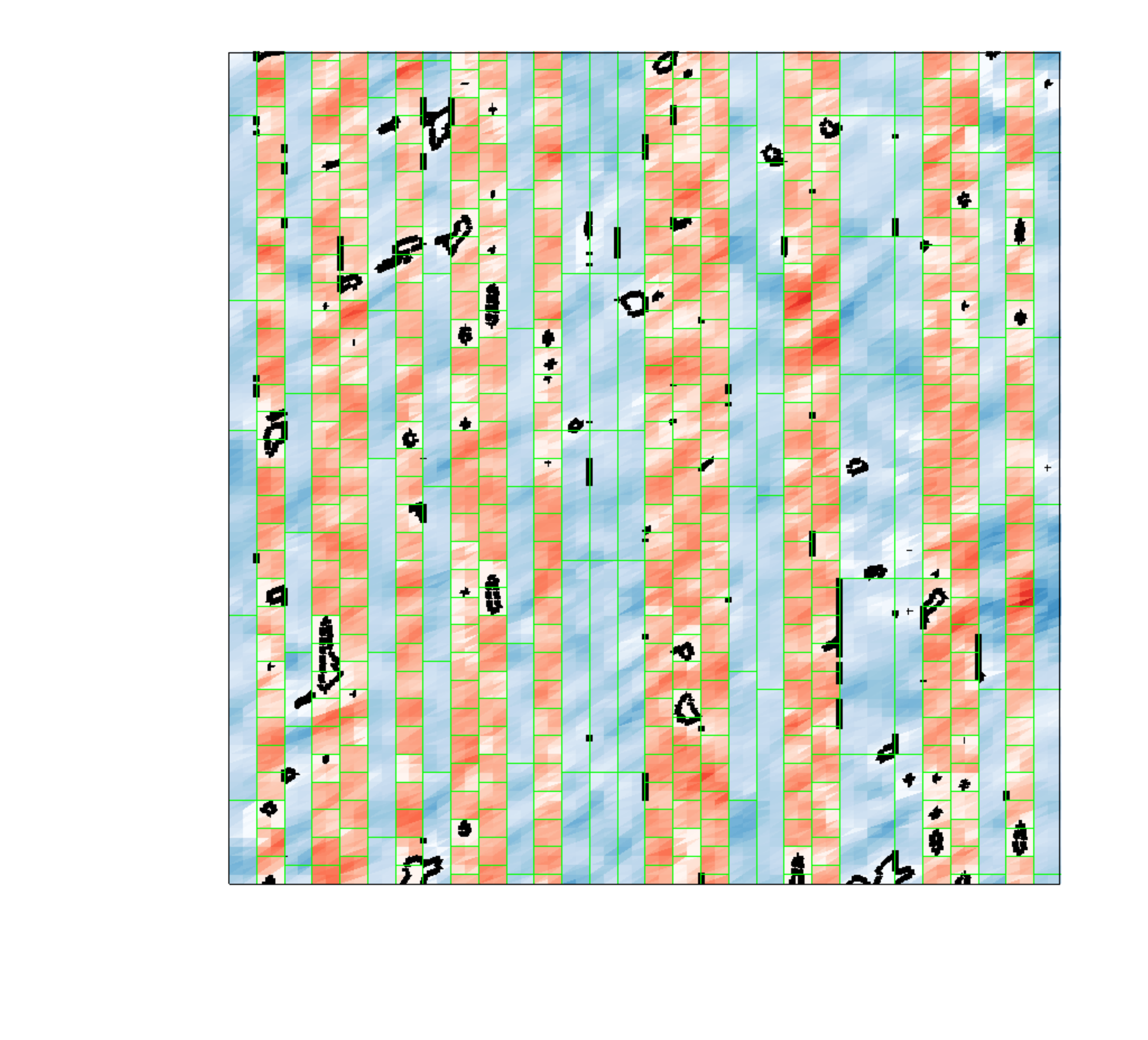}}
\subfigure[$x=0.875$]{\includegraphics[trim = 2.5in 1.5in 0.9in 0.5in, clip, width = 0.30\textwidth]{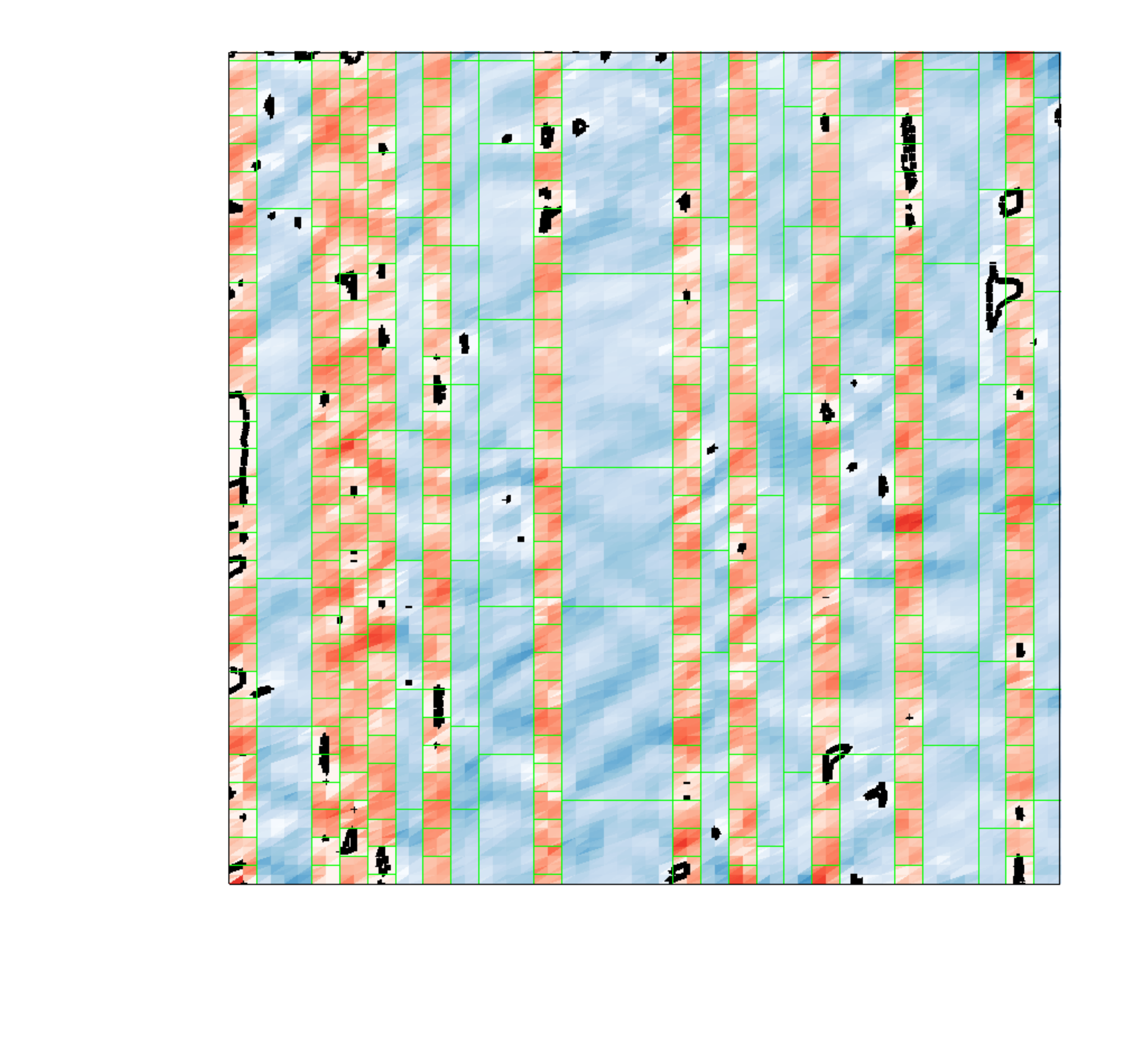}}
\subfigure[$x=1.000$]{\includegraphics[trim = 2.5in 1.5in 0.9in 0.5in, clip, width = 0.30\textwidth]{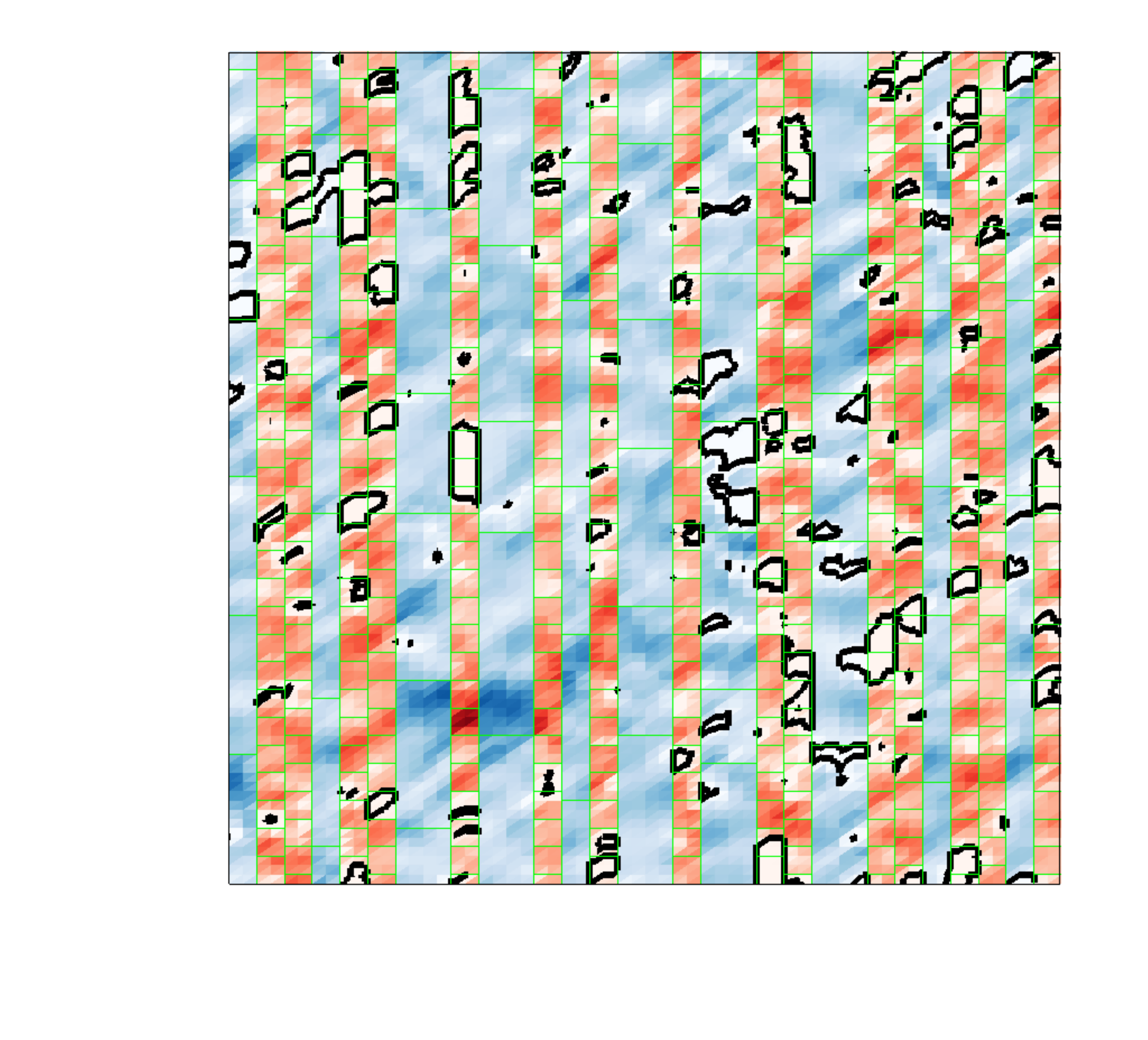}}
\subfigure{\includegraphics{DpeffLegend}}
\caption{Effective plastic deformation rate at macroscopic yield for uniaxial loading. Contours corresponding to $D^p_{\mathit{eff}} = 10^{-5}~\mathrm{s}^{-1}$ are plotted. This threshold value corresponds to one-tenth of the macroscopic deformation rate.}
\label{fig:DPeff2D}
\end{figure}

Views of an LDX-2101 aggregate at yield under uniaxial loading are presented in Figures~\ref{fig:DPeff3D} and~\ref{fig:DPeff2D}. The plots depict the effective plastic deformation rate $D^p_{\mathit{eff}}$, which serves as an indicator for yield. These plots were constructed using plastic deformation rate data from an elasto-plastic finite element simulation. Elements whose effective plastic deformation rate is at least one-tenth of the effective macroscopic deformation rate are considered yielded. Contours depicting the boundary between elastic and plastic regions are shown. Figure~\ref{fig:DPeff3D} presents a three-dimensional view and Figure~\ref{fig:DPeff2D} presents several two-dimensional cross-sections along the $x$-axis. 
The strain rates vary spatially by a factor of three to four.  The variations appear as 
clear diagonal bands running from the  lower left to upper right.  Bands also run from  lower right to upper left, but are not quite as evident.  The bands cross both phases and extend through the interior.

\begin{figure}[ht]
\centering
\includegraphics[trim = 1in 1in 1in 1.5in, clip, width = 0.7\textwidth]{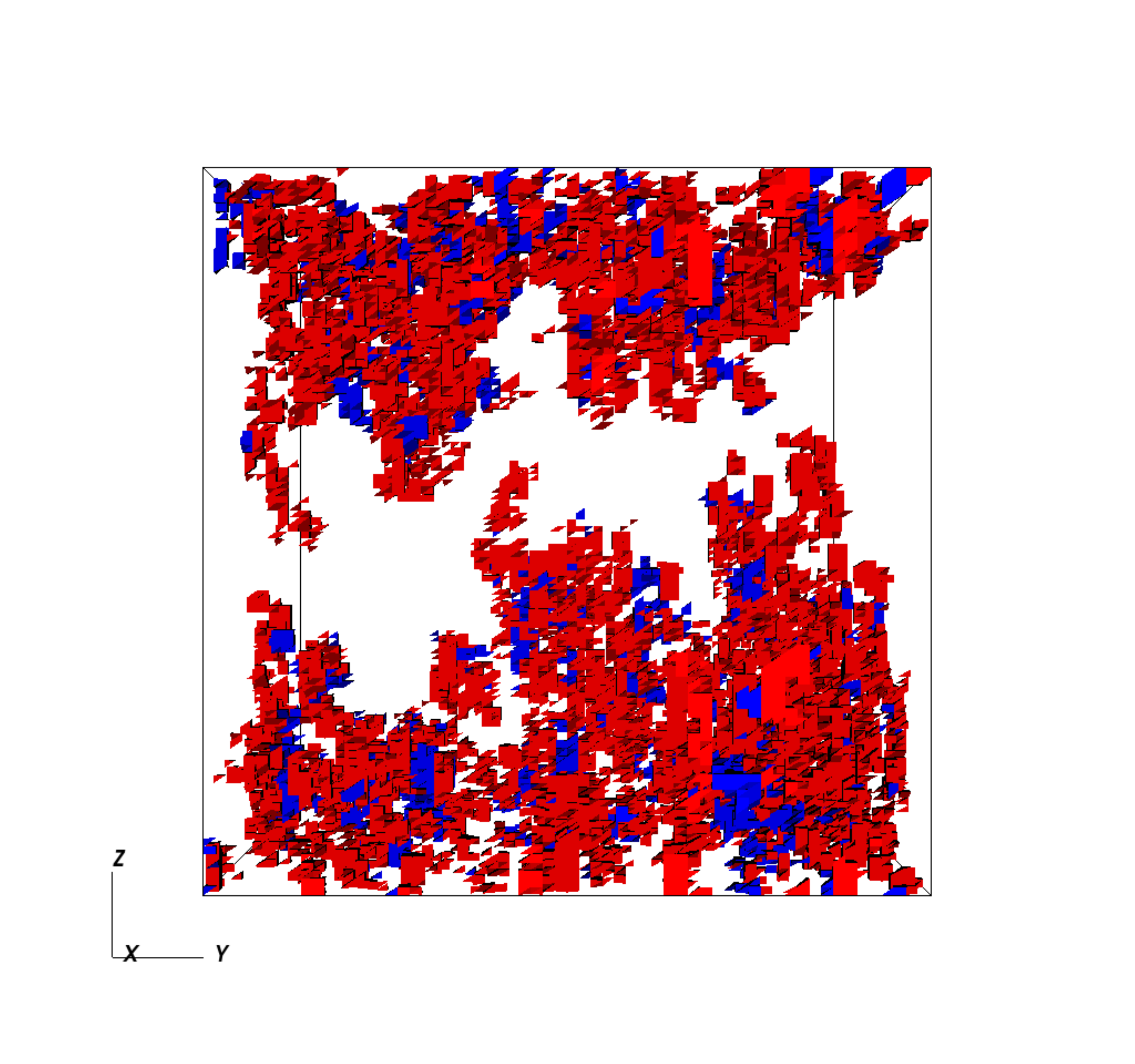}
\caption{Anatomy of a yield band. Only elastic elements connected to the top and bottom mesh surfaces are shown. Austenite is red. Ferrite is blue.}
\label{fig:YieldBand}
\end{figure}

The yield band detected for the uniaxial loading deformation field shown in is shown in Figures~\ref{fig:DPeff3D} and~\ref{fig:DPeff2D} appears in Figure~\ref{fig:YieldBand}. 
In this figure, austenite is red, and ferrite is blue. On can readily see the a complete gap between elastic 
elements attaced to the top surface and those attached to the bottom surface. 
Only elastic elements connected to the top and bottom mesh surfaces are shown. 
The two surfaces are separated by the yield band. 
The yield band consists of plastic regions and islands of elastic material completely surrounded by plastic regions. 
The band runs diagonally across the same, consistent with the presence of bands of
higher plastic strain rate shown in Figures~\ref{fig:DPeff3D} and \ref{fig:DPeff2D}.

For LDX-2101 with the microstructure described, the formation of a yield band first occurs when 92\% of the aggregate has yielded.   This is less than the critical volume fraction of 96\% for a random distribution of yielding. It is expected that late in the elasto-plastic transition the relative load rate would increase dramatically for the critical areas maintaining the connection between two surfaces. As a result, these critical areas yield preferentially. The yielded volume fraction at macroscopic yield is therefore less than the critical volume fraction for a random distribution of yielding.
\clearpage
\subsection{Yield surface for biaxial stress states}
\label{sec:yield_surf}
\begin{figure}[ht]
\centering
\includegraphics[trim = 0.1in 0.4in 0.2in 0.2in, clip, width = 0.7\textwidth]{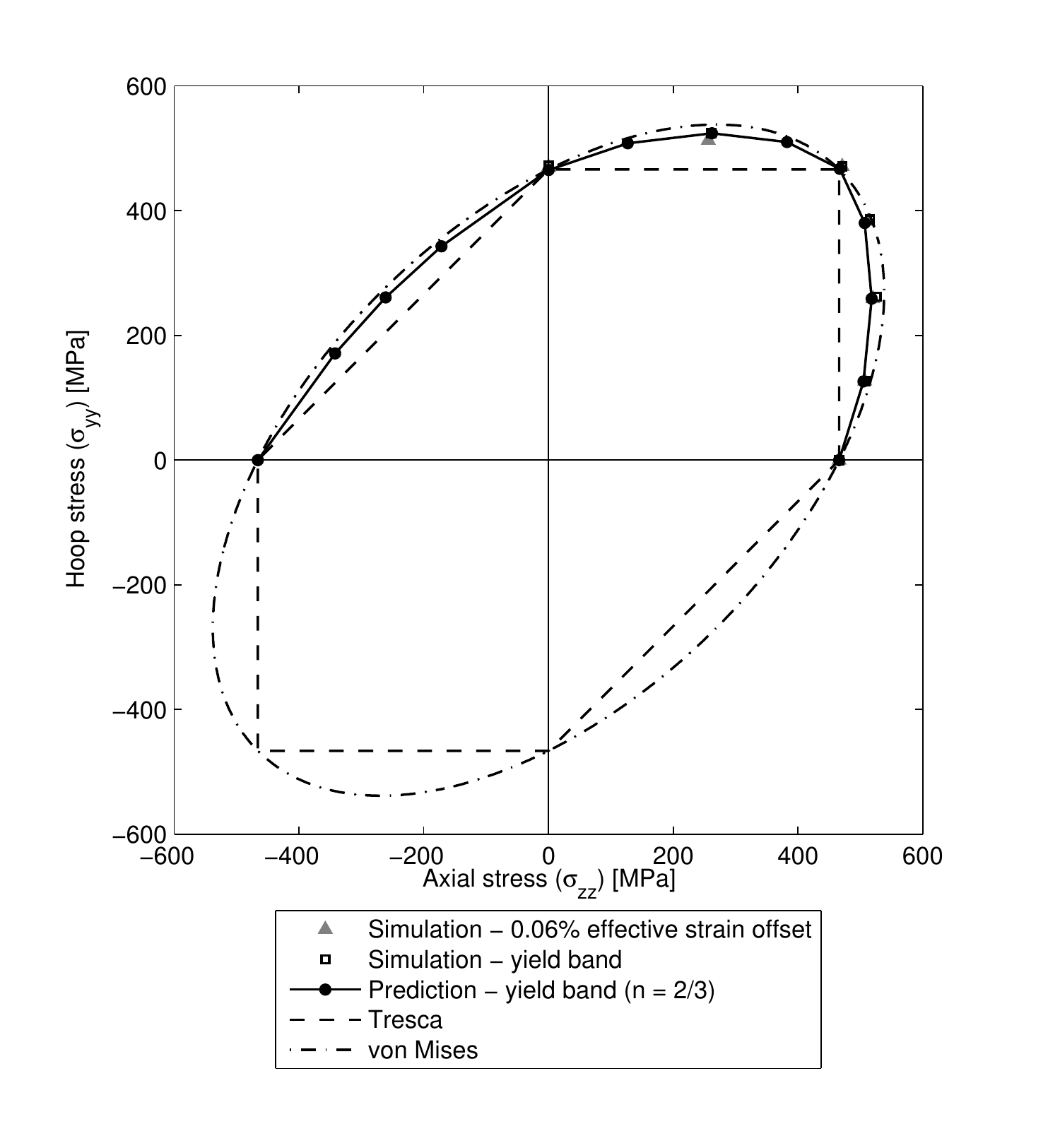}
\caption{Simulated and predicted yield surfaces for biaxial loading.}
\label{fig:YieldSurface}
\end{figure}

The plane stress yield surface was constructed from the results for all the stress states probed and  is shown in Figure~\ref{fig:YieldSurface}.  
The specific stress states analyze served as nodal points for a piecewise representation of the
complete surface, as described in Section~\ref{sec:piecewise-reps}.
Von Mises and Tresca surfaces also are shown and can be seen to bound the predicted surface.
For comparison, the yield surface was also evaluated  for five points on the yield surface using the full elasto-plastic simulation with the yield band detection algorithm. 
The values obtained from this techniques also are shown in Figure~\ref{fig:YieldSurface}.
In addition to the graphical depection shown in Figure~\ref{fig:YieldSurface}, the axial components of the macroscopic yield stresses are tabulated in Table~\ref{tab:AxialYieldStress}. The predicted and simulated yield stresses, both computed using the yield band detection algorithm, exhibit good agreement, within 2\% of each other. 
The yield band detection algorithm, coupled with the predictive method for determining elemental macroscopic yield stresses, is therefore an accurate and computationally efficient method for evaluating macroscopic yield stress. 
\begin{table}[ht]
	\centering
	\caption{Axial component of yield stress for LDX-2101 in MPa.}	
	\begin{tabular} {c c c c}
		 BR & Prediction & Simulation & Simulation \\
		 & yield band & yield band & strain offset \\
		 & ($\pm1$ MPa) & ($\pm2$ MPa) & \\ \hline
		 0.00 & 466 & 466 & 466 \\
		 0.25 & 504 & 508 & 507 \\
		 0.50 & 518 & 526 & 520 \\
		 0.75 & 506 & 515 & 512 \\
		 1.00 & 466 & 471 & 471
	\end{tabular}
	\label{tab:AxialYieldStress}
\end{table}

\begin{figure}[ht]
\centering
\includegraphics[width = 0.5\textwidth]{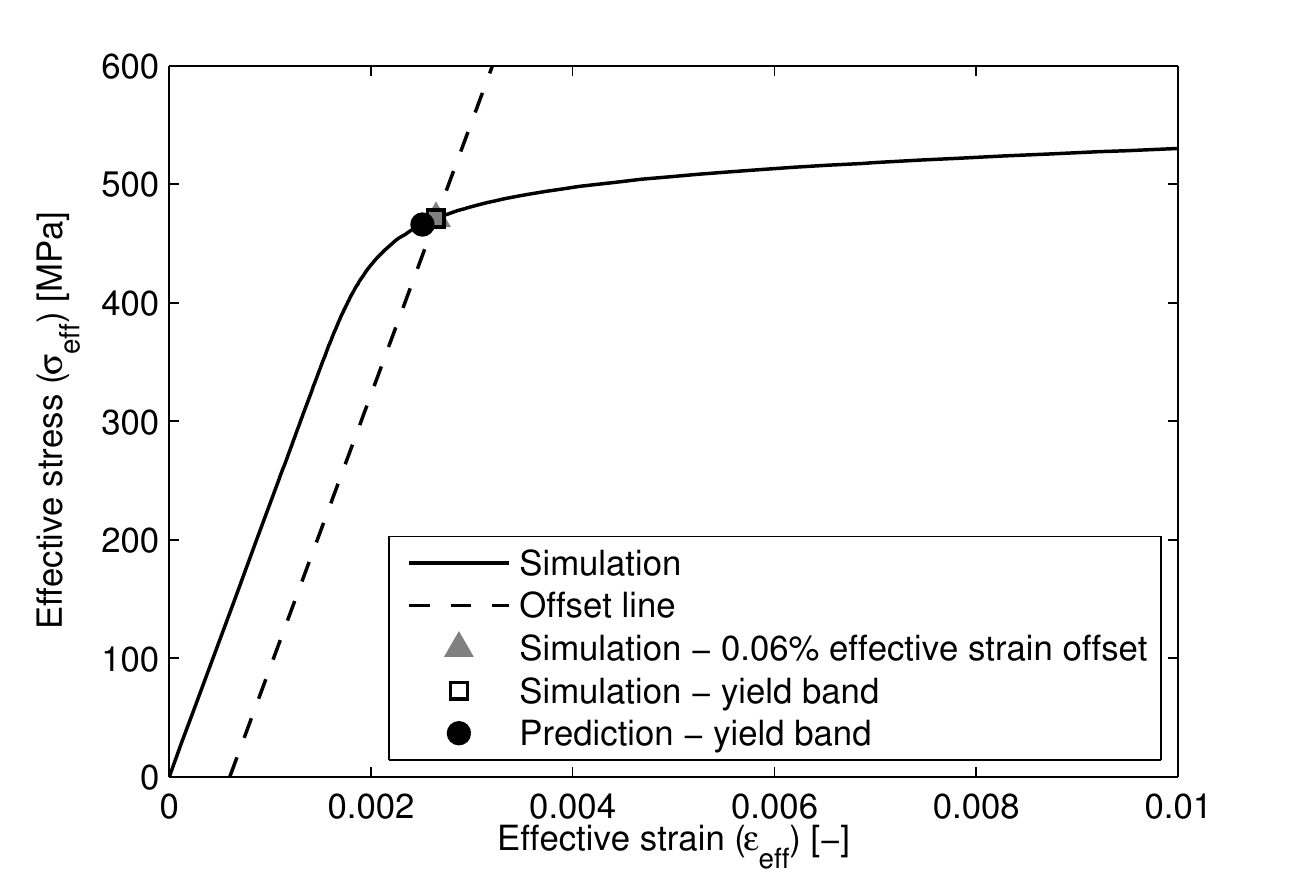}
\caption{Effective stress-strain curve for LDX-2101 at a biaxial ratio of unity. One measure of effective yield stress is given by the intersection of the effective stress-strain curve and a line drawn parallel to the elastic region at a prescribed offset. In this case, an offset of 0.06\% strain was chosen to produce good agreement with the yield band detection algorithm. The components of the yield stress tensor are determined from the effective yield stress and the direction of loading in deviatoric stress space.}
\label{fig:EffStressStrainBR100}
\end{figure}

The predicted yield stresses also agree well with those computed using the 0.06\% effective strain offset method, which is illustrated in Figure~\ref{fig:EffStressStrainBR100}. 
This approach uses  the traditional strain offset method~\cite{Rees06a} from the elastic portion of the stress-strain curve.
The offset of 0.06\% was chosen to match the yield surface data, which highlights the fact that the size of the offset is arbitrary. The yield stresses determined using the offset method are dependent on the size of the offset. The advantage of the yield band formulation is that it is physically based, and does not rely on an arbitrary offset. If an offset method is to be used, the yield band detection formulation can inform the size of the offset. In this example, the macroscopic material strength is fairly isotropic in the plane of loading considered, as reflected by the similarity of the predicted and von Mises yield surfaces. The new yield detection algorithm is, however, generally applicable to materials with highly anisotropic macroscopic strength.

To demonstrate the effect of relative phase strengths on macroscopic yield surface, the yield surface was evaluated using the predictive methodology for an unequal phase strength aggregate. The initial slip system strengths ($g_0$) were chosen such that the initial slip system strength for ferrite was twice that of austenite. The volume-average slip system strength is the same as for the equal phase strength case. The saturation slip system strenghts ($g_s$) were also adjusted such that the two phases hardened at the same rate. The saturation slip system strengths do not, however, affect the yield prediction, and are included only for completeness of the material description. The material parameters are tabulated in Table~\ref{tab:PlasticityParams}. The same material microstructure was used for both the equal and unequal phase strength cases. The only difference between the two cases is the slip system strengths.

\begin{figure}[ht]
\centering
\includegraphics[trim = 0.1in 0.4in 0.2in 0.2in, clip, width = 0.7\textwidth]{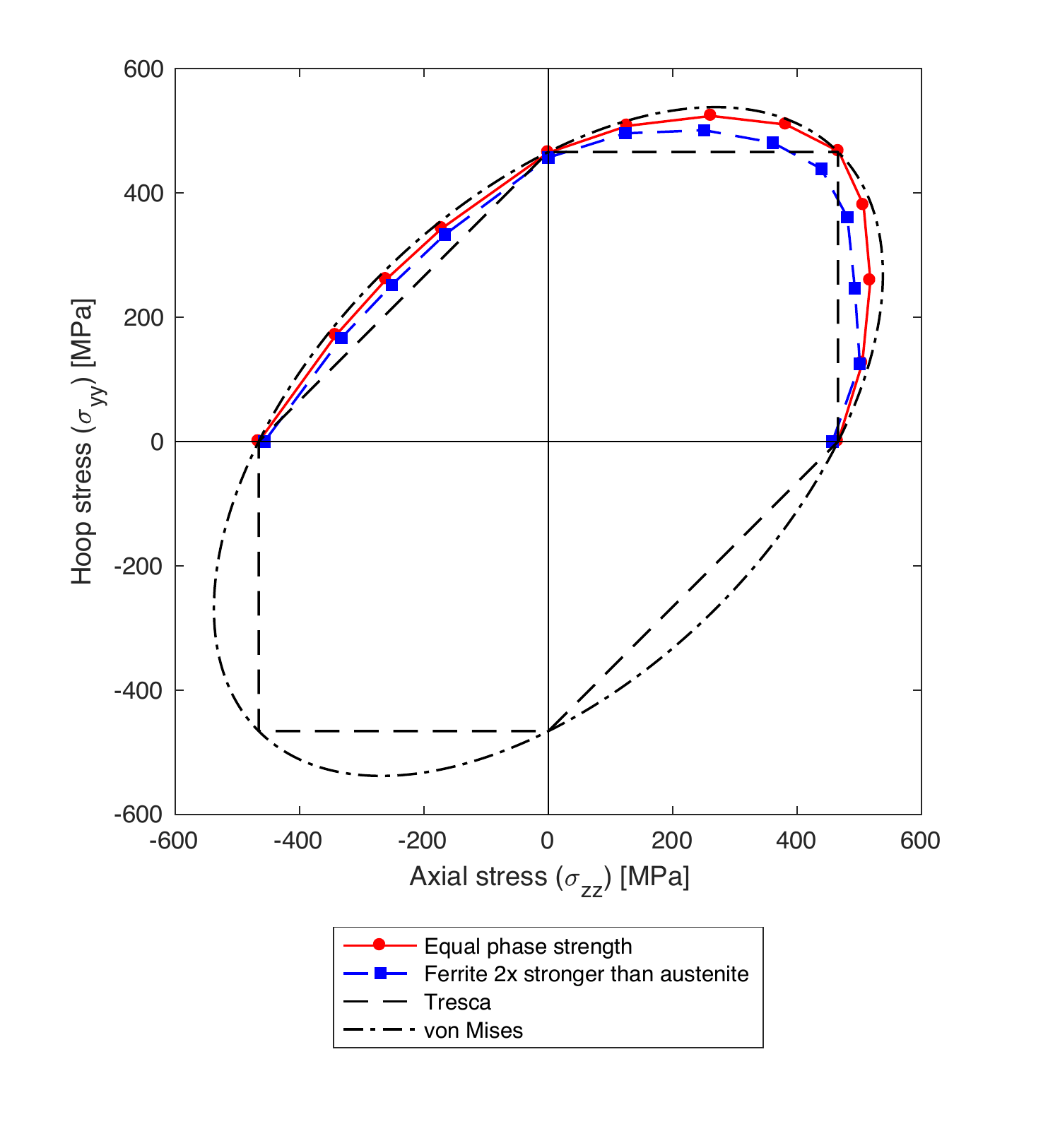}
\caption{Simulated and predicted yield surfaces for biaxial loading for equal and unequal phase strength cases. The prefential yielding of austenitic regions for the unequal phase strength case, coupled with the phase topology, leads to a reduction in yield strength for the unequal phase strength case under biaxial loading.}
\label{fig:YieldSurface_twocases}
\end{figure}

The predicted yield surfaces for the equal and unequal phase strength cases are presented in Figure~\ref{fig:YieldSurface_twocases}. Changing the relative strength of the two phases alters the topology of the yield surface. The macroscopic yield strength is same for both cases for uniaxial loading in both the axial and transverse (hoop) directions. A decrease in macroscopic yield strength is observed for the unequal phase strength material for all nonzero values of biaxial ratio, with the greatest decrease in strength occuring for balanced biaxial loading.

For the unequal phase strength case, the initial slip system strength of austenite is half that of ferrite. The strength-to-stiffness ratios for austenitic regions are, in general, lower than the strength-to-stiffness ratios for ferritic regions. As a result, one would expect the majority of austnitic regions to yield before the majority of ferritic regions. Since the austenitic regions form a continuous banded structure running through the aggregate, complete yeilding of the austenitic regions corresponds to the formation of a yield band. Thus, a lower volume fraction of yielded regions is required for macroscopic yielding for the unequal phase strength case, owing to the phase topology of the microstructure. This behavior is observed in the predictive model. Macroscopic yielding occurs when 90-93\% of the equal phase strength aggregate is yielded locally. In contrast, macroscopic yielding occurs when only 62-77\% of the unequal phase strength aggregate is yeilded locally. The preferential yielding of austenitic regions that connect to form a yield band accounts for the reduction in strength of the unequal phase strength aggregate.
\clearpage
\section{Discussion}
\label{sec:discussion}

 \subsection{Usefulness of the new methodology}
 The most direct application of the method is generate a macroscopic yield surface from microstructural characterization data.  This is particularly useful if available experimental testing capabilities are limited to a few loading modes, such as tension/compression or tension/torsion, but service conditions for the material might involve more general loading conditions.  In such cases, which are common in engineering applications, combining limited experimental data with estimates from the new method could be a very effective approach to estimating the yield surface in a timely manner.  Presently, existing forms of the yield condition are invoked, like the von Mises criterion for isotropic behaviors or the Barlat criterion for anisotropic behaviors under plane stress states (as reviewed in Section~\ref{sec:background}).  The new method extends these approaches to include the explicit incorporation of microstructural state.  
 
Another major use for the new method is expected to be investigations of the sensitivity of microstructural features 
on the macroscopic performance of a material.  Multiphase alloys in particular often exhibit a wide array of
microstructural states depending on the processing history.  Processing variables associated with forming schedules 
and heat treatments affect the morphology and topology of the phases.  These, in turn, affect such attributes as the phase contiguity, which is known to strongly influence strength in dual phase systems having large differences in phase strength, for example.  In addition, the phases that comprise a dual or multiphase alloy usually are not pure elements
even though the composition may be dominated by one element.  Mechanical properties for crystals of the phase composition may differ from those of a pure material.  Thus, two general topics of interest for sensivity  studies are:
\begin{itemize}
    \item Phase properties: single crystal values of elastic moduli and slip system strength both change the strength-to-stiffness metric and thus change the spatial distribution of local yielding as the load is increased.
    This, in turn, influences the heterogeneity of deformation, the onset of a macroscopic yield band, and the macroscopic yield surface.   As shown in the LDX-2101 application here, these factors influence the shape of the
    yield surface and thus the degree of anisotropy.  An interesting attribute of the new method is that changes in slip system strengths do not affect the directional stifffness, so examining strength variations can be done readily as only one set of finite element simulations are needed to quantify the directional stiffness.  
    \item Microstructural state: phase volume fractions, phase topology, grain shape and lattice orientation all factor into the manner that phases and grains mechanically interact.  Again, this influences the heterogeneity of the deformation and the onset of local yielding. Spatial variations in the yield band follow from the spatial variations in the constituent phases and grains, leading to differences in the yield surface with variations microstructure.   Examination of the impact of changing the  thermomechanical processing, and thus changing the microstructre,  are open to study using the new method. 

\end{itemize}

 \subsection{Issues concering virtual samples}
 Central to the method are virtual samples.  The  method relies on the virtual samples for (1) accurate evaluation of directional stiffness and (2) replication of critical features of the microstructure to enable locating  yield bands.  These are two aspects of creating what are commonly referred to in the literature as 
 representative volume elements (RVE).   By representative, it is meant that any one sample will produce a response that 
falls within a typical distribution of observed behaviors  and that the responses of a collection of samples will reproduce the observed distribution.   A few comments are offered here regarding issues related to the two
primary functions of the samples (quantifying directional stiffness and locating yield bands).  These are discussed in terms of two types of microstructures:  non-periodic and periodic.  
 
\begin{figure}[ht]
\centering
\includegraphics[width = 0.6\textwidth]{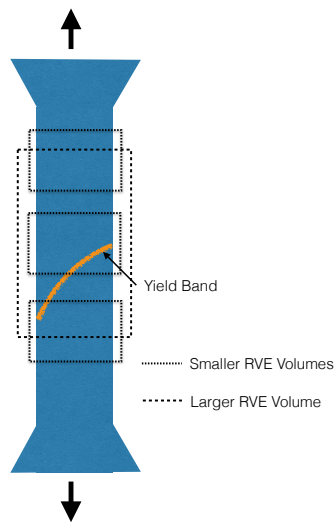}
\caption{Schematic of tensile sample having a yield band that extends across the gage section and showing possible RVE volumes extracted from the gage section.   }
\label{fig:rve-tensile}
\end{figure}

{\it Samples with non-periodic microstructures.}
  Related work on two-phase titanium samples~\cite{ti_rve_2018}, shown schematically in Figure~\ref{fig:rve-tensile}, indicate that RVE's taken from the gage section of one sample (with defined spatial arrangement of grains) gave lower  strength as the RVE size was increased.  This happens because a smaller RVE may not coincide with the location of the initial band, but instead bases the onset of yielding on a band within the RVE that occurs at a higher load when local yielding is spatially more extensive.  The full sample always gives the lowest strength.  To accurately assess the macroscopic strength the sample must be sufficiently large to capture the yield band regardless of the orientation it prefers withing the sample.  The microstructure influences the orientation of the yield band with the effect that the RVE size depends on the sample microstructure.  This was observed for samples drawn from textured plate (parallel versus perpendicular to the rolling direction of the plate).  Thus, caution is needed to assure that the sample is capable of exhibiting the controlling macroscopic deformation mode for yielding.  

\begin{figure}[ht]
\centering
\includegraphics[width = 0.5\textwidth,angle = 90]{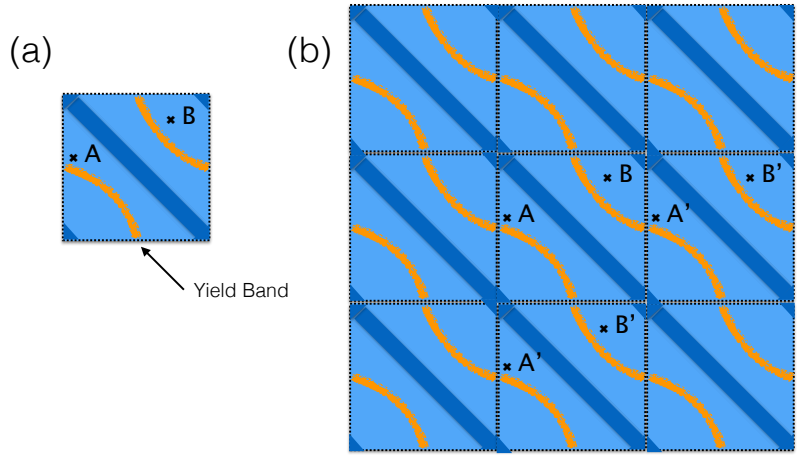}
\caption{Material with layered microstructure (lighter and darker blue bands) with yield bands indicated by orange lines.  (a) Unit cell of microstructure. (b) virtual sample created by tiling with the unit cell. A and B designate finite elements of the discretized sample. }
\label{fig:symmetry_bc}
\end{figure}

{\it Samples with periodic microstructures.} 
Highly repetive microstructures can present an interesting challenge in terms of detecting a yield band.   If the virtual sample is constructed to exploit periodicity of the microstructure (perhaps by assuming a more perfect arrangement of phases and grains than actually exists), then a constraint may arise from implementing the method using samples tied to a Cartesion domain.  An example is offered in Figure~\ref{fig:symmetry_bc}, in which a repetive microstructure shown (b) is built from the unit cell shown in (a). If the method were applied to a sample containing  only the unit cell shown in (a) and not invoking periodicity, macroscopic yielding would not be detected for the yield bands shown because the domain is artificially truncated.  If, on the other hand, meshes with periodic microstructure and periodic boundary conditions were used with the method,  the yield criterion could be that the material is yielded when there does not exist any element in the mesh that can be connected to all of its images via a flood-fill algorithm. Looking at Figure~\ref{fig:symmetry_bc}-b, neither Element A nor B (nor any other element for that matter) can be connected to all of its images (A' and B' respectively). Applying the detection algorithm in this way would indicate that the material is yielded. This approach would eliminate any dependence on the shape of the domain. 

\section{Conclusions}
\label{sec:conclusions}
A new method to determine macroscropic yield surfaces is presented that explicitly includes
the material's microstructure in construction of the yield surface.  
The method is physically-based in that it links the distribution of yielding
at the microscale to macroscopic yielding. The method predicts the onset of local yielding
and then detects the existence of a macroscopic yield band, which is defined as an interconnected
region of yielded material that completely separates two opposing
surfaces of a polycrystalline aggregate. Central to the new method is a correlation between
the onset of yielding locally and a metric for the crystal-scale, multiaxial strength-to-stiffness ratio.  
Existence of a yield band spanning the sample can be detected using either an eigenmode analysis or a flood-fill algorithm. 

The new yield method is used to determine the yield surface for a dual-phase stainless steel 
for plane stress loading conditions.  It is compared to the yield surface found by conducting
elastoplastic simulations through the full elastoplastic transition to estimate the onset of 
macroscopic yielding.   The yield surface predicted using the new methodology shows good agreement
with the surface estimated from the full simulations.  
Further,  agreement is good with the simulated yield surface calculated using the traditional strain offset method. 

The yield surface also is evaluated for an aggregate with unequal phase strengths. Even though the volume-averaged slip system strengths were the same for both the equal and unequal phase strength aggregates, the unequal phase strength aggregate demonstrated a reduction in strength under biaxial loading. The decrease in strength and the alteration of the yield surface topology are attributed to the preferential yielding of austenitic regions that interconnect to form a yield band.

The new method provides an efficient method for evaluating a macroscopic yield surface while accounting 
for the grain and phase neighborhood effects associated with microstructure.  The data needed to utilize the new method consists of the  strain distribution for only a single elastic loading step
for each of the loading modes, plus knowledge of the single crystal slip system strengths.
It is advantageous over the traditional strain offset method because it is 
based on the progation of local yielding through the microstructure, rather than the size of the
arbitrary offset in the macroscopic stress-strain response.

\section{Acknowledgements}
\label{sec:acknowledgements}
Support was provided  by the US Office of Naval Research (ONR) under contract N00014-09-1-0447.
\clearpage

\bibliographystyle{unsrt}
\bibliography{References}
\clearpage

\appendix
\section{Yield detection using eigenmode analyses} 
\label{sec:appendix_a}

To detect the existence of a yield band for multiaxial loading, a purely elastic auxiliary system is considered. In this auxiliary system, elastic regions have nonzero stiffness and plastic regions have zero stiffness. Essential boundary conditions of zero normal velocity are imposed on each of the coordinate planes. The global stiffness matrix is assembled and its nullspace computed. The nullspace of the stiffness matrix is comprised of all deformations that produce zero strain energy. Any deformation in the nullspace can be represented as a linear combination of orthonormal basis vectors that span the nullspace. The dimension of the nullspace is equal to the number of zero eigenvalues of the stiffness matrix. Trial displacement fields are used to test for yielding. Each trial displacement field represents uniform displacement along one of the coordinate directions. Each trial displacement field is projected, in turn, onto the nullspace. If, for any trial displacement field, the projected surface displacements match the trial surface displacements, it can be concluded that the surface is separated from the opposing surface by a yield band, which implies macroscopic yield. Otherwise, the material is still macroscopically elastic.

The analysis applies to a rectangular prismatic material domain consisting of both elastic and plastic regions. As an example, the case of uniaxial loading, in which the load is applied to two opposing surfaces via rigid platens, is first considered.

\begin{figure}[ht]
\centering
\subfigure[Undeformed]{\includegraphics[width = 0.32\textwidth, trim = 0.4in 0.5in 0.4in 0.5in, clip]{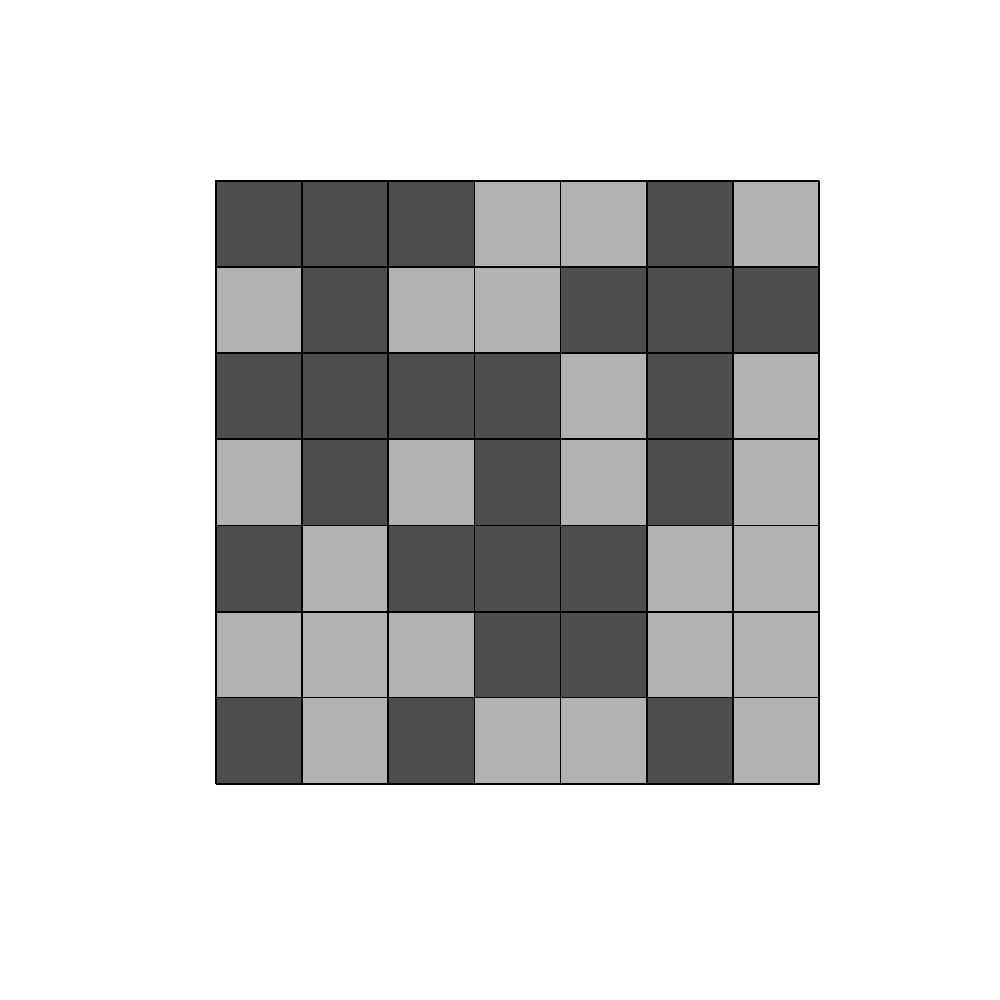}}
\subfigure[$x$-displacement]{\includegraphics[width = 0.32\textwidth, trim = 0.4in 0.5in 0.4in 0.5in, clip]{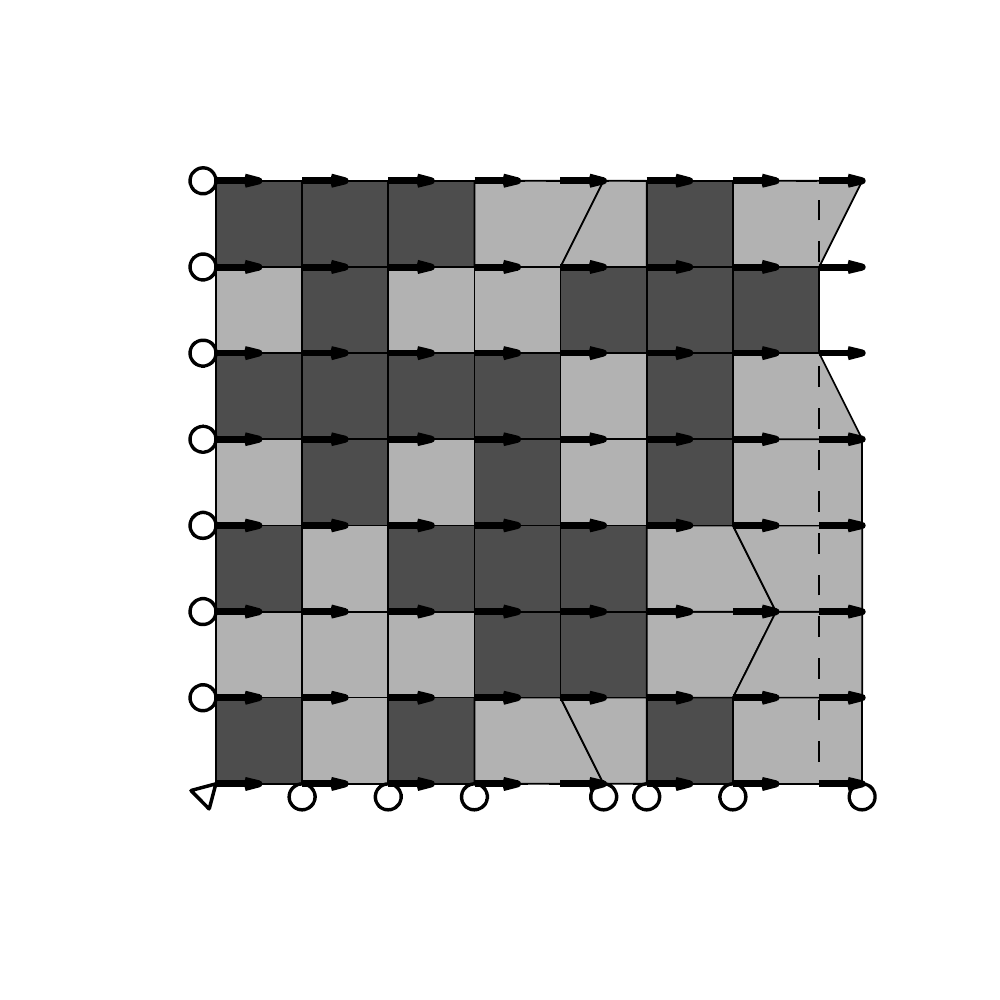}}
\subfigure[$y$-displacement]{\includegraphics[width = 0.32\textwidth, trim = 0.4in 0.5in 0.4in 0.5in, clip]{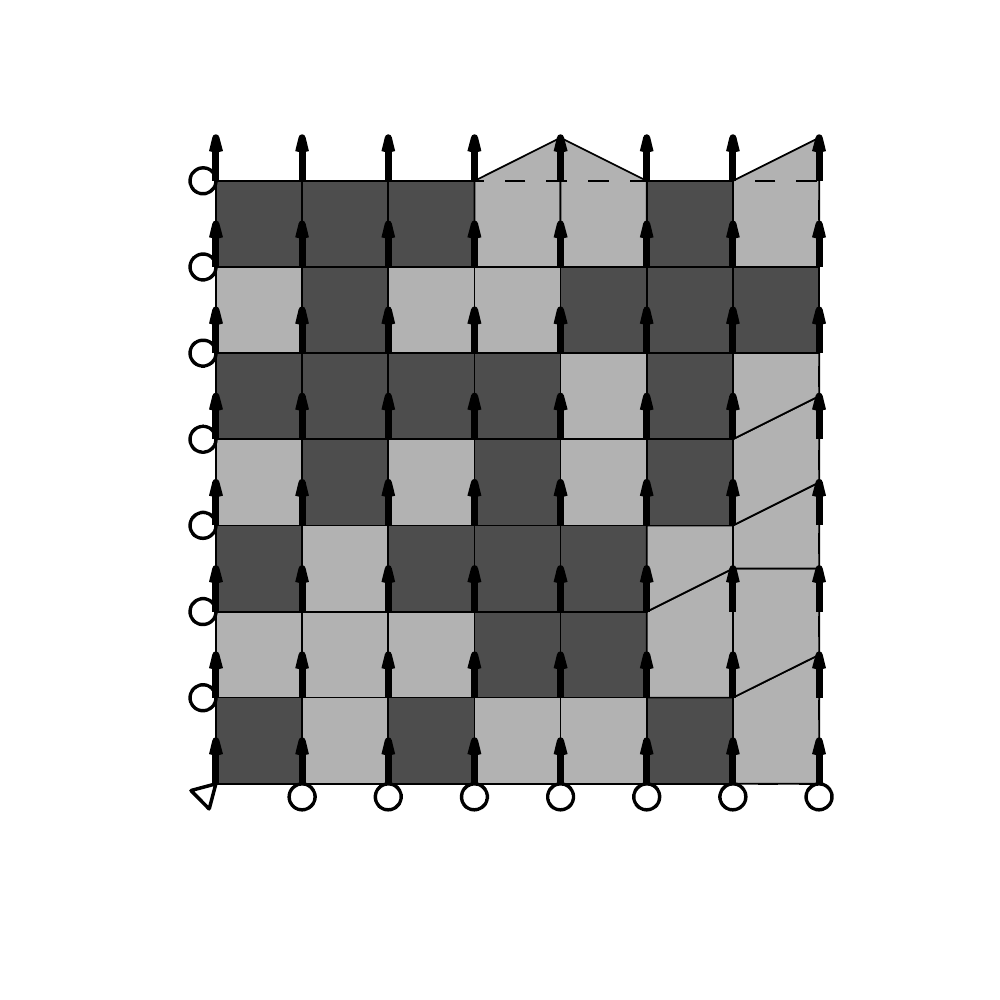}}
\caption{Macroscopically elastic material.}\label{fig:2D_yield_ex1}
\end{figure}

\begin{figure}[ht]
\centering
\subfigure[Undeformed]{\includegraphics[width = 0.32\textwidth, trim = 0.4in 0.5in 0.4in 0.5in, clip]{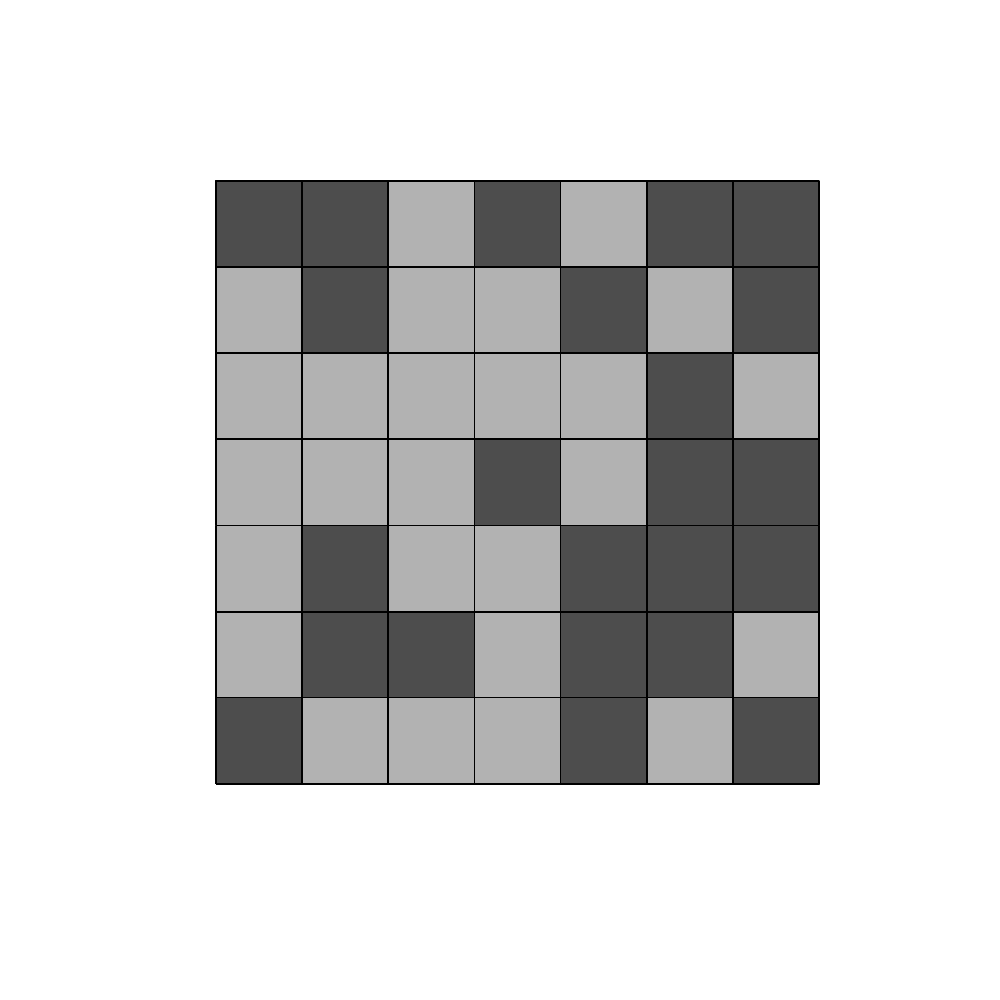}}
\subfigure[$x$-displacement]{\includegraphics[width = 0.32\textwidth, trim = 0.4in 0.5in 0.4in 0.5in, clip]{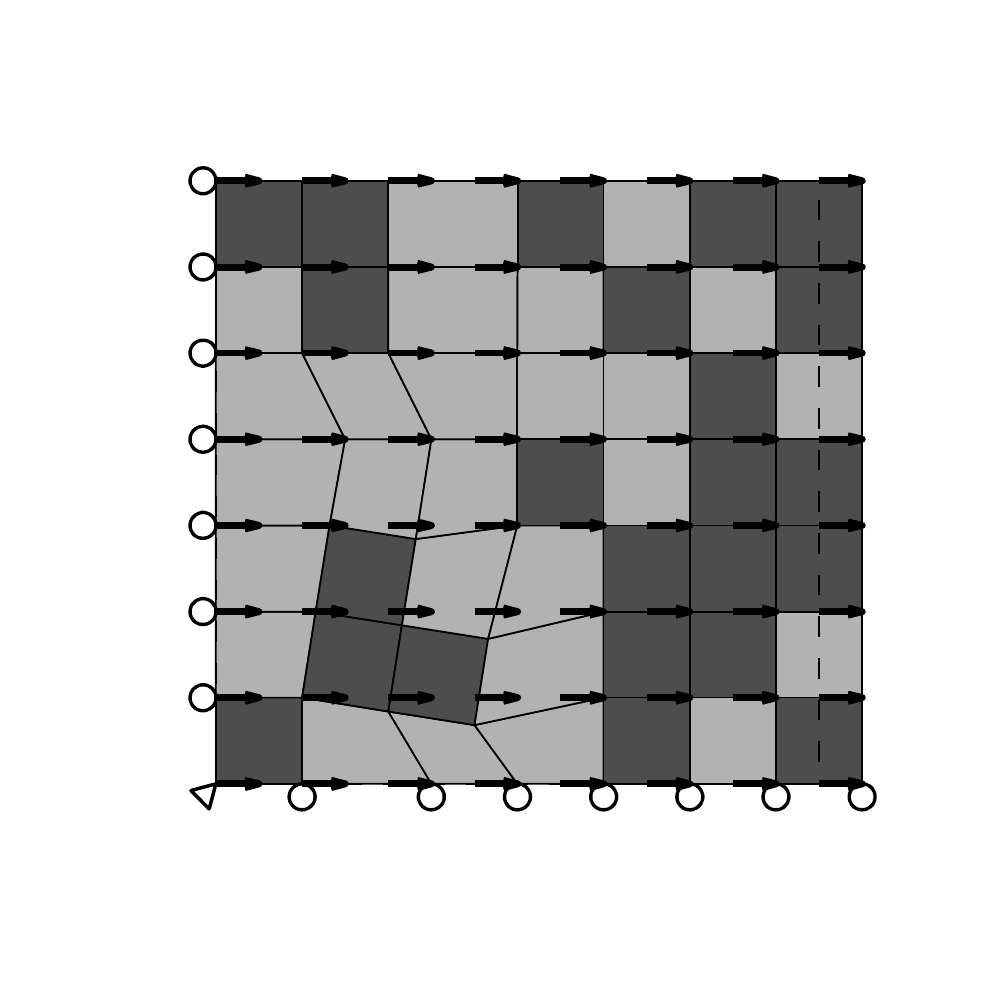}}
\subfigure[$y$-displacement]{\includegraphics[width = 0.32\textwidth, trim = 0.4in 0.5in 0.4in 0.5in, clip]{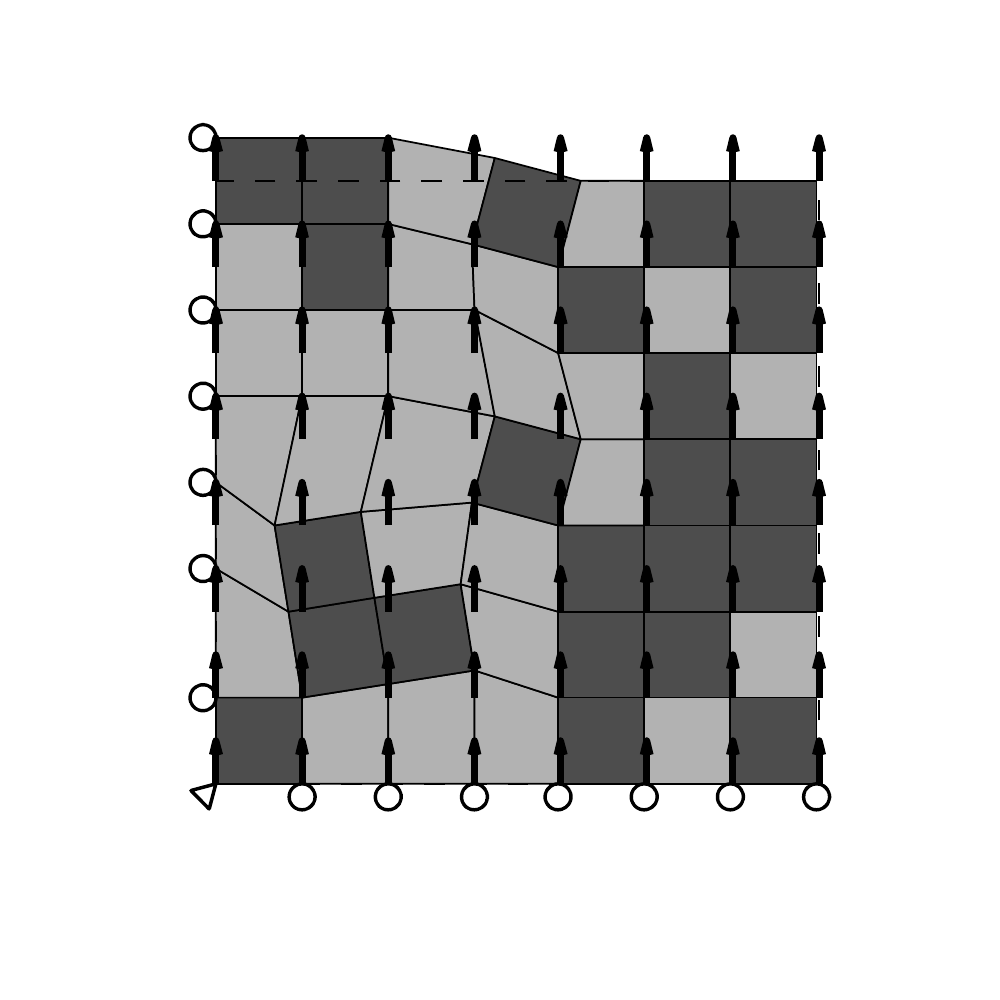}}
\caption{Vertical yield band.}\label{fig:2D_yield_ex2}
\end{figure}

\begin{figure}[ht]
\centering
\subfigure[Undeformed]{\includegraphics[width = 0.32\textwidth, trim = 0.4in 0.5in 0.4in 0.5in, clip]{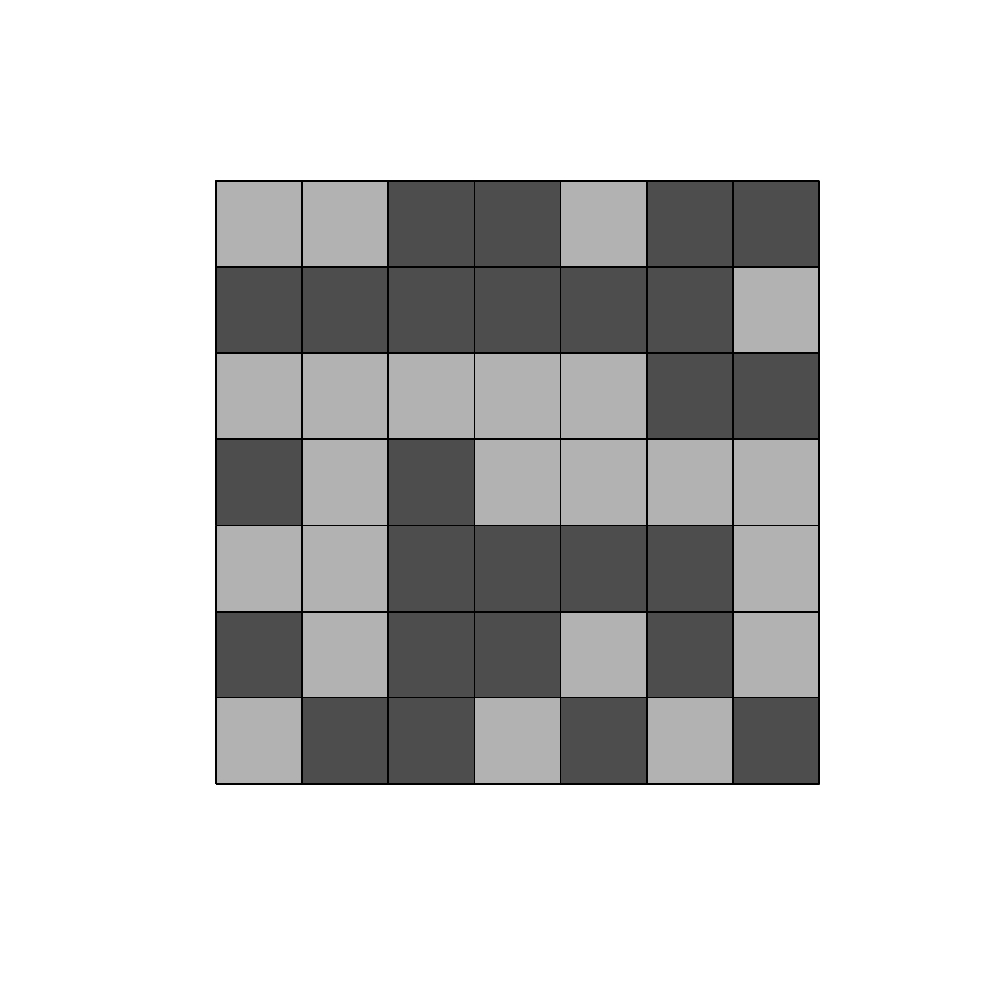}}
\subfigure[$x$-displacement]{\includegraphics[width = 0.32\textwidth, trim = 0.4in 0.5in 0.4in 0.5in, clip]{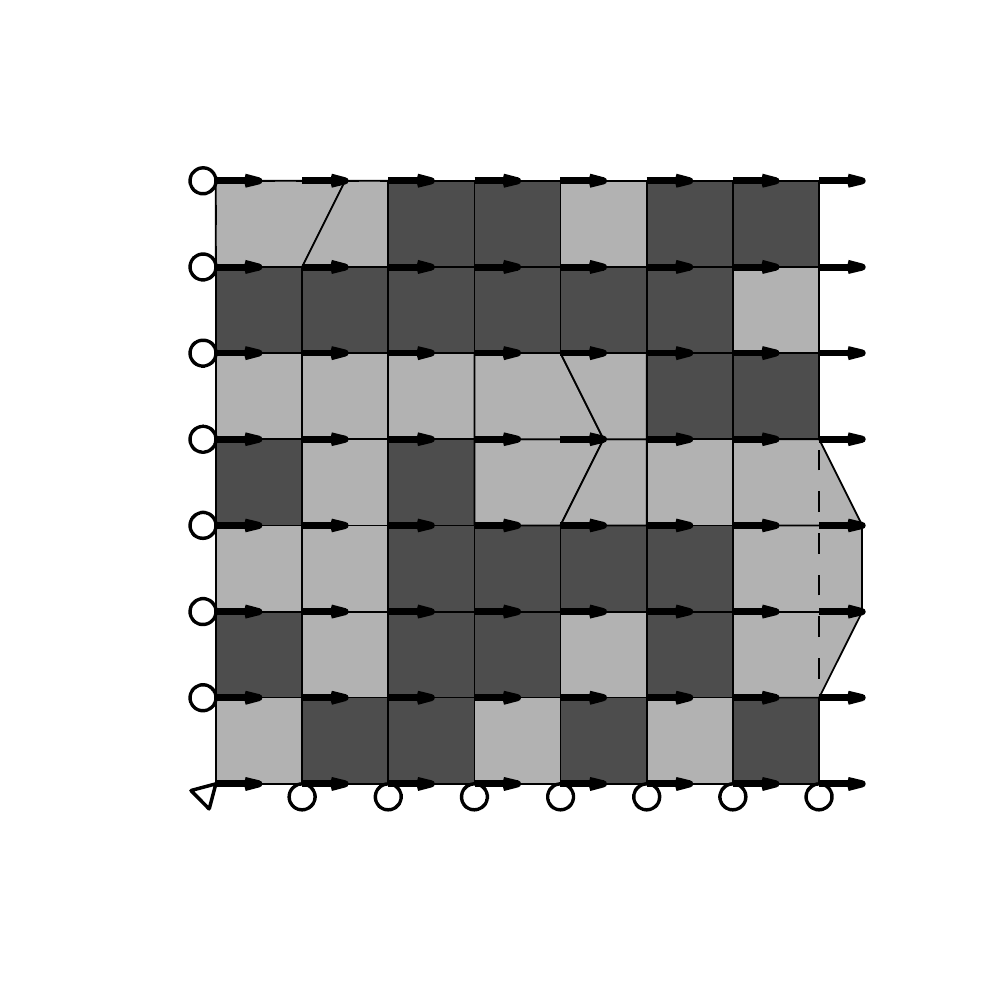}}
\subfigure[$y$-displacement]{\includegraphics[width = 0.32\textwidth, trim = 0.4in 0.5in 0.4in 0.5in, clip]{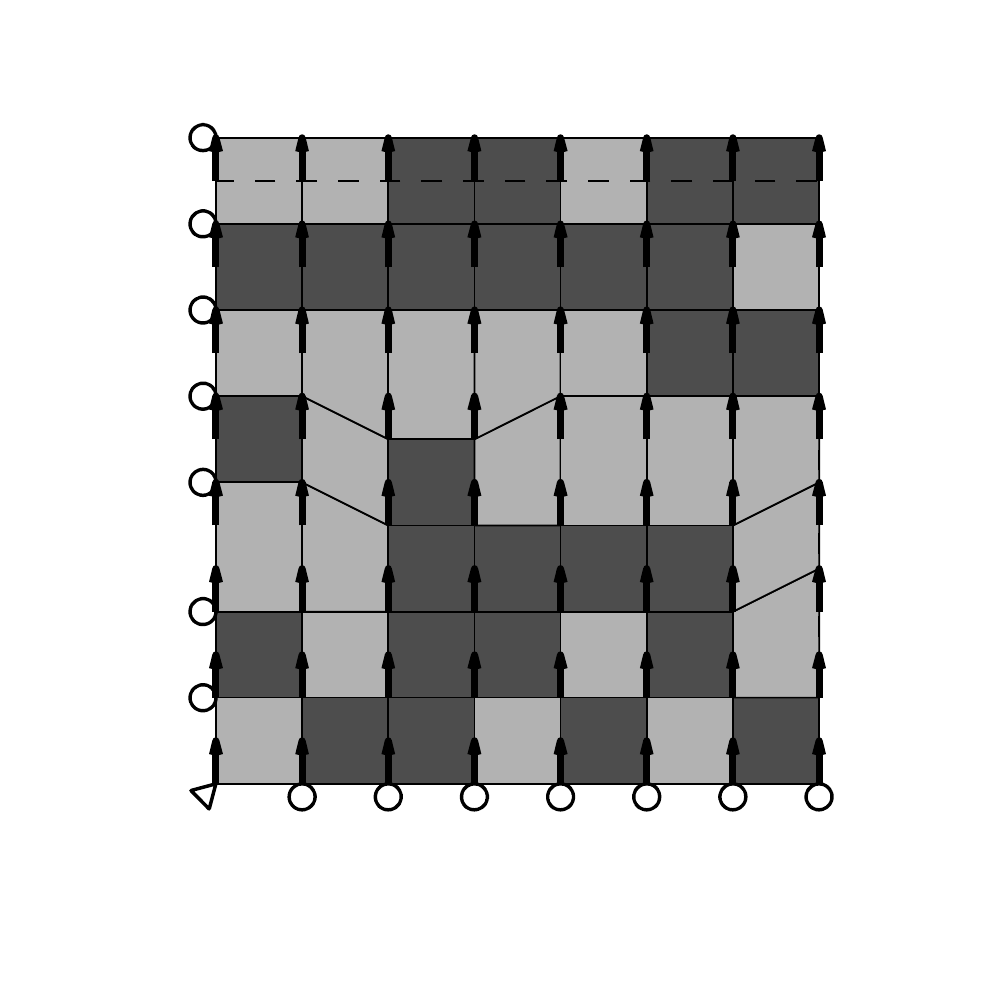}}
\caption{Horizontal yield band.}\label{fig:2D_yield_ex3}
\end{figure}

\begin{figure}[ht]
\centering
\subfigure[Undeformed]{\includegraphics[width = 0.32\textwidth, trim = 0.4in 0.5in 0.4in 0.5in, clip]{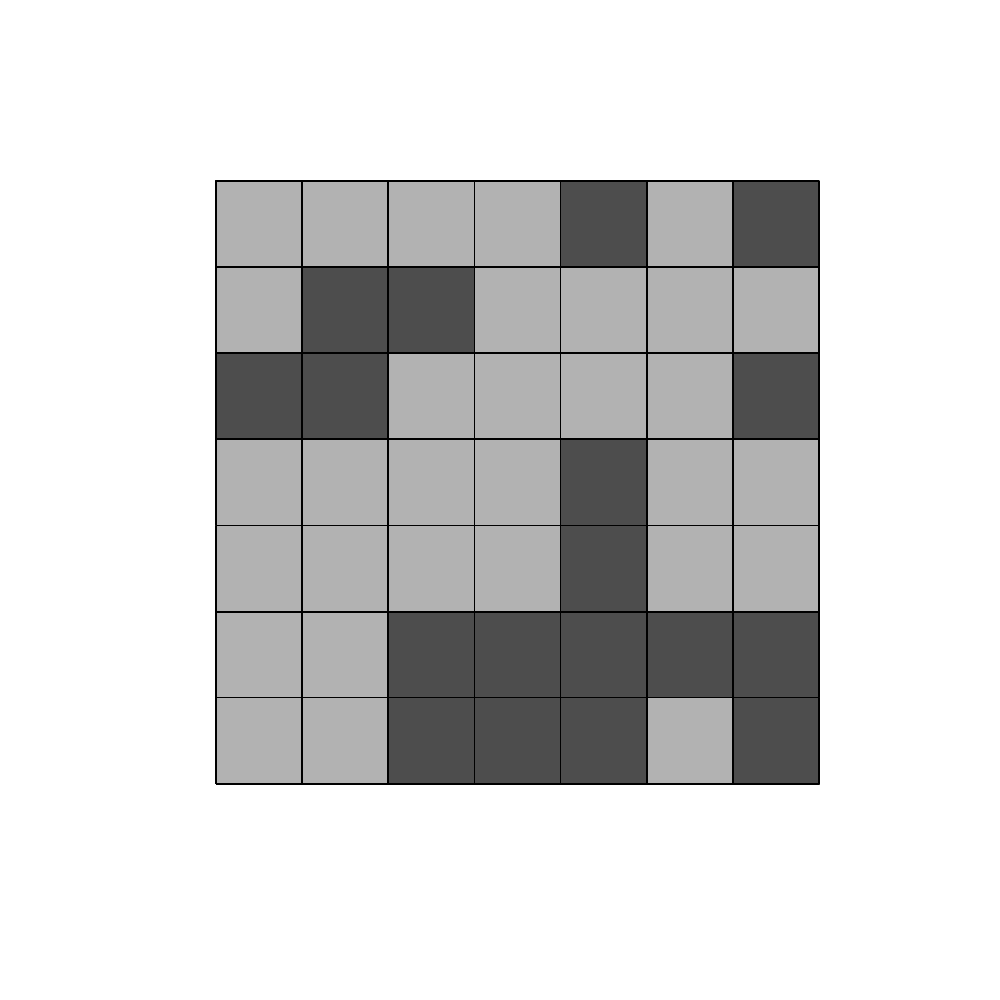}}
\subfigure[$x$-displacement]{\includegraphics[width = 0.32\textwidth, trim = 0.4in 0.5in 0.4in 0.5in, clip]{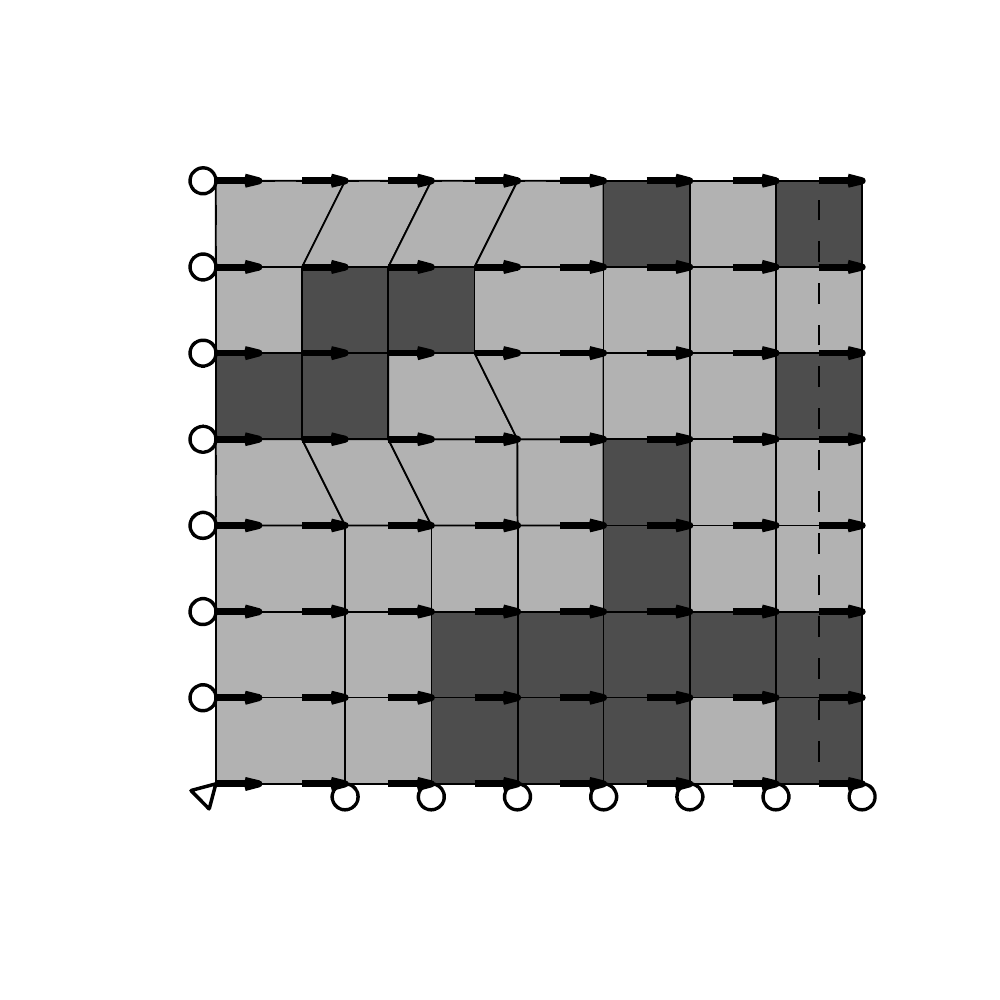}}
\subfigure[$y$-displacement]{\includegraphics[width = 0.32\textwidth, trim = 0.4in 0.5in 0.4in 0.5in, clip]{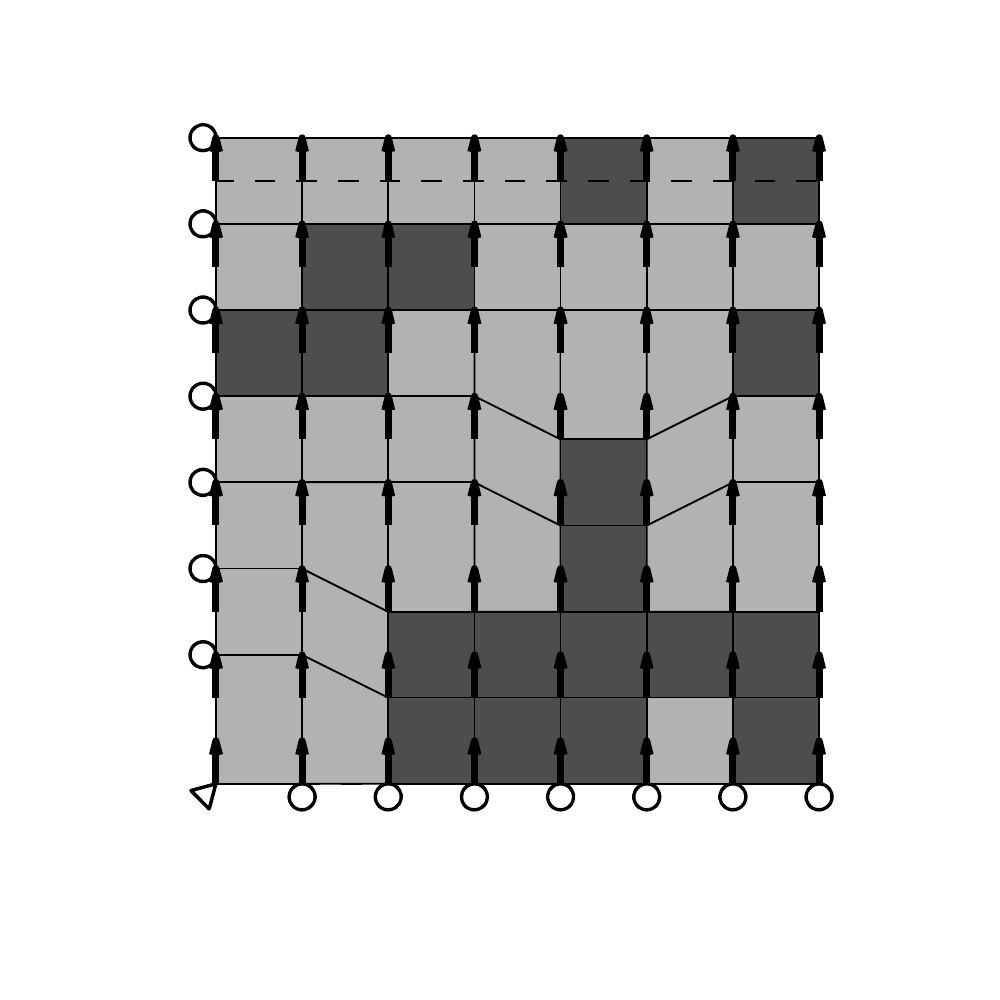}}
\caption{Vertical and horizontal yield bands.}\label{fig:2D_yield_ex4}
\end{figure}

Two-dimensional examples are presented in Figures~\ref{fig:2D_yield_ex1}-\ref{fig:2D_yield_ex4}. In these examples, dark gray represents elastic regions and light gray represents plastic regions. The dashed outline demarcates the boundary of the undeformed mesh. The boundary conditions and trial displacement fields applied to the auxiliary system are shown. The trial displacement magnitude is half an element width. In the first example, (Figure~\ref{fig:2D_yield_ex1}), there is no yield band. Both surfaces translate non-uniformly and the material is therefore elastic. In the second example (Figure~\ref{fig:2D_yield_ex2}), there is a yield band running vertically through the material. The right surface translates uniformly by the trial displacement and the material is therefore yielded. The third example in Figure~\ref{fig:2D_yield_ex3} is similar to the second, except that the yield band runs horizontally through the material and the top surface translates uniformly. In the fourth example (Figure~\ref{fig:2D_yield_ex4}) there are both horizontal and vertical yield bands. Both surfaces translate uniformly by the trial displacement, and the material is yielded.

\clearpage

\section{Crystal plasticity constitutive equations} 
\label{sec:appendix_b}
The constitutive model employs a kinematic decomposition of the deformation into a sequence of deformations due to crystallographic slip, rotation, and elastic stretch. Using this decomposition, the deformation gradient, $\boldsymbol{F}$, can be represented as
\begin{equation}\label{eqn:kinematic_decomp}
\boldsymbol{F} = \boldsymbol{V}^e  \boldsymbol{R}^*  \boldsymbol{F}^p 
\end{equation}
where $\boldsymbol{F}^p$, $\boldsymbol{R}^*$, and  $\boldsymbol{V}^e$ correspond to crystallographic slip, rotation, and elastic stretch, respectively.  This decomposition defines a reference configuration $\mathscr{B}_0$, a deformed configuration $\mathscr{B}$, and two intermediate configurations $\bar{\mathscr{B}}$ and $\hat{\mathscr{B}}$. The state equations are written in the intermediate $\hat{\mathscr{B}}$ configuration defined by the relaxation of the elastic deformation from the current $\mathscr{B}$ configuration. Elastic strains, $\boldsymbol{\epsilon}^e$, are required to be small, which allows the elastic stretch tensor to be written as
\begin{equation}
\boldsymbol{V}^e = \boldsymbol{I} + \boldsymbol{\epsilon}^e
\end{equation}
The velocity gradient, $\boldsymbol{L}$, is calculated from the deformation gradient using the relationship
\begin{equation}
\boldsymbol{L} = \dot{\boldsymbol{F}} \boldsymbol{F}^{-1}
\end{equation}
The velocity gradient can be decomposed into a symmetric deformation rate, $\boldsymbol{D}$, and skew-symmetric spin rate, $\boldsymbol{W}$
\begin{equation}
\boldsymbol{L} = \boldsymbol{D} + \boldsymbol{W}
\end{equation}
A generic symmetric tensor, $\boldsymbol{A}$, can be additively decomposed into mean and deviatoric components
\begin{equation}\label{eqn:vol_dev_decomp}
\boldsymbol{A} = \sfrac{1}{3} \mathrm{tr} (\boldsymbol{A}) \boldsymbol{I} + \boldsymbol{A}^\prime
\end{equation}
where $\sfrac{1}{3} \mathrm{tr} (\boldsymbol{A})$ is the scalar mean component and $\boldsymbol{A}^\prime$ is the tensorial deviatoric component.
Utilizing Equations~\ref{eqn:kinematic_decomp}-\ref{eqn:vol_dev_decomp} the volumetric deformation rate, deviatoric deformation rate, and spin rate can be expressed as
\begin{equation}
\mathrm{tr} \left( \boldsymbol{D} \right) = \mathrm{tr} \left( \dot{\boldsymbol{\epsilon}}^e \right)
\end{equation}
\begin{equation}
\boldsymbol{D}^\prime = \dot{\boldsymbol{\epsilon}}^{e\prime} + \hat{\boldsymbol{D}}^p + \boldsymbol{\epsilon}^{e\prime} \cdot \hat{\boldsymbol{W}}^p - \hat{\boldsymbol{W}}^p \cdot \boldsymbol{\epsilon}^{e\prime}
\end{equation}
\begin{equation}\label{eqn:spin_rate}
\boldsymbol{W} = \hat{\boldsymbol{W}}^p + \boldsymbol{\epsilon}^{e\prime} \cdot \hat{\boldsymbol{D}}^p - \hat{\boldsymbol{D}}^p \cdot \boldsymbol{\epsilon}^{e\prime}
\end{equation}
where $\hat{\boldsymbol{D}}^p$ and $\hat{\boldsymbol{W}}^p$ are the plastic deformation and spin rates.

Constitutive equations relate the stress to the deformation. The Kirchhoff stress, $\boldsymbol{\tau}$, in the $\hat{\mathscr{B}}$ configuration is related to the Cauchy stress, $\boldsymbol{\sigma}$, in the current configuration $\mathscr{B}$ by the determinant of the elastic stretch tensor
\begin{equation}
\boldsymbol{\tau} = \mathrm{det} (\boldsymbol{V^e}) \boldsymbol{\sigma}
\end{equation}
The Kirchhoff stress is related to the elastic strain through anisotropic Hooke's law
\begin{equation}
\mathrm{tr} \left( \boldsymbol{\tau} \right) = 3K \mathrm{tr} \left( \boldsymbol{\epsilon}^e \right)
\end{equation}
\begin{equation}
\boldsymbol{\tau}^\prime = \mathscr{C}^* \boldsymbol{\epsilon}^{e\prime}
\end{equation}
where $K$ is the bulk modulus and $\mathscr{C}^*$ is the fourth-order stiffness tensor.

Plastic deformation due to crystallographic slip occurs on a restricted set of slip systems. For FCC crystals, slip occurs on the \{111\} planes in the [110] directions. For BCC crystals, slip on the \{110\} planes in the [111] directions is considered. The plastic deformation rate and plastic spin rate are given by 
\begin{equation}
\hat{\boldsymbol{D}}^p = \sum_\alpha \dot{\gamma}^\alpha \hat{\boldsymbol{P}}^\alpha
\end{equation}
\begin{equation}
\hat{\boldsymbol{W}}^p = \dot{\boldsymbol{R}^*} \boldsymbol{R}^{*T} + \sum_\alpha \dot{\gamma}^\alpha \hat{\boldsymbol{Q}}^\alpha
\end{equation}
where $\hat{\boldsymbol{P}}^\alpha$ and $\hat{\boldsymbol{Q}}^\alpha$ are the symmetric and skew symmetric components of the Schmid tensor $\hat{\boldsymbol{T}}^\alpha$, and $\dot{\gamma}^\alpha$ is the shear rate on the $\alpha$-slip system. The Schmid tensor is defined as the dyad of the slip direction, $\hat{\mathbf{s}}^\alpha$, and slip plane normal, $\hat{\mathbf{m}}^\alpha$
\begin{equation}
\hat{\boldsymbol{T}}^\alpha = \hat{\mathbf{s}}^\alpha \otimes \hat{\mathbf{m}}^\alpha
\end{equation}
The slip system shear rate for a given slip system is related to the critical resolved shear stress on that slip system, $\tau^\alpha$, by a power law relationship
\begin{equation}
\dot{\gamma}^\alpha = \dot{\gamma}_0 \left( \frac{\vert \tau^\alpha \vert}{g^\alpha} \right)^\frac{1}{m} \mathrm{sgn} \left( \tau^\alpha \right)
\end{equation}
where $\dot{\gamma}_0$ is a reference slip system shear rate and $g^\alpha$ is the slip system strength. The resolved shear stress is the projection of the deviatoric stress onto the slip system
\begin{equation}
\tau^\alpha = \boldsymbol{\tau}^\prime : \hat{\boldsymbol{P}}^\alpha
\end{equation}

A modified Voce hardening law is used to describe slip system strength evolution 
\begin{equation} \label{eqn:Voce}
\dot{g}^\alpha = h_0 \left( \frac{g_s - g^\alpha}{g_s - g_0} \right)^{n^\prime} \sum_\alpha \dot{\gamma}^\alpha
\end{equation}
where $h_0$ is the reference hardening rate, $g_s$ is the saturation strength, and $n^\prime$ is the hardening exponent. The hardening law is isotropic; at a given material point, all slip systems harden at the same rate. Equation~\ref{eqn:spin_rate} describes the crystal reorientation.

\end{document}